# Title: Behavioral change models for infectious disease transmission: a systematic review (2020–2025)

**Authors:** Youngji Jo[1]*, Sileshi Sintayehu Sharbayta[1], Bruno Buonomo[2]

**Affiliations**

1. Department of Public Health Sciences, School of Medicine, University of Connecticut, Farmington, Connecticut, USA
2. Department of Mathematics and Applications, University of Naples Federico II, I-80126 Naples, Italy; buonomo@unina.it

**Corresponding author**

Youngji Jo: jo@uchc.edu

**Abstract**

**Background:** Human behavior fundamentally shapes infectious disease dynamics, yet integrating behavioral processes into transmission models remains inconsistent and theoretically fragmented. Recent epidemics—especially COVID-19—have intensified the need for mathematical models that capture how individuals and communities adapt to perceived risk, social influence, and policy signals. This systematic review synthesizes post-2020 infectious disease models that incorporate behavioral adaptation mechanisms, characterizes their theoretical grounding, and evaluates how behavioral constructs modify transmission, vaccination, and compliance outcomes.

**Methods:** Following PRISMA guidelines, we searched Scopus and PubMed (2020–2025) and screened 1,274 unique records, supplemented by citation chaining of key review papers. We extracted standardized data on disease context, country, modeling framework, behavioral adaptation mechanisms (prevalence-dependent, policy/media, imitation/social learning), and psychosocial constructs (personal threat, coping appraisal, barriers, social norms, cues to action), informed by major behavior theories. A total of 216 studies met the inclusion criteria.

**Results:** COVID-19 dominated the literature (73%), with most models using compartmental ordinary differential equation frameworks (81%) and theoretical or U.S.-based populations. Behavioral change was primarily reactive: 47% used prevalence-dependent feedback, where distancing or care-seeking intensified with rising cases or deaths; 25% incorporated information/media-based awareness dynamics; and 19% relied on exogenous policy triggers such as lockdowns. Game-theoretic or social learning mechanisms were rare (≤5%). Behavioral changes overwhelmingly targeted the contact/transmission rate (91%). Psychosocial constructs were unevenly represented: cues to action (n=159) and personal threat (n=145) dominated, while coping appraisal (n=82), barriers (n=36), and social norms (n=25) appeared far less often.

**Conclusions:** We propose a unifying taxonomy organized around behavioral drivers, social scale, and memory structure to highlight dominant modeling paradigms, evaluate their empirical grounding, and recommend pathways for more rigorous integration of behavioral science into epidemiological modeling. By mapping models to established psychosocial constructs, our approach offers a novel framework for guiding future behavior-adaptation model formulation, comparative evaluation, and data collection strategies.

## 1 Introduction

Social and behavioral factors are critical to the emergence, spread and containment of human disease and act as key determinants of the course, duration and outcomes of outbreaks.[1,2-4] Recent large-scale epidemics, such as COVID-19, have reinforced the need for models that better integrate social and behavioral dynamics and theories to reflect the realities of affected communities.[5,6] Individuals and communities play a vital role in reducing transmission during outbreak response by maintaining preventive behaviors.[7,8] The feasibility and acceptability of adopting recommended health behaviors are influenced by complex elements, including the individual's perception of the disease threat, their level of trust in governing authorities and other information sources, and their physical, financial, and social capacity to voluntarily take action.[7,9] Ultimately, incorporating these factors into disease models is expected to improve their predictive accuracy and enable the design of more effective response measures and policies.

Modeling approaches for integrated behavior often augment classical epidemiological frameworks with mechanisms to capture dynamic behavioral shifts, which is defined as behavior changing endogenously as a function of another time-dependent variable within the model.[10] In a review of COVID-19 models incorporating adaptive (endogenous) behavior, three main approaches were identified: feedback loops (72% of studies), game theory/utility theory (27%), and information/opinion spread (9%).[11] The feedback loop approach uses the prevalence of a disease outcome (such as cases or deaths) to stimulate a change in behavior within the model, sometimes leading to the projection of periodic waves of infection.[6] Game-theoretic methods analyze strategic interactions by assuming rational decision-makers who weigh the costs and benefits or maximize utility.[12] In these models, individuals reduce contacts to balance the benefits of social interaction against the perceived risk and costs associated with contracting the disease. Finally, models focused on information/opinion spread (or "coupled contagion") simulate how an individual's actions are influenced by the transmission of attitudes or awareness through social networks or social media.[13] These processes capture how opinions—which may or may not reflect the objective epidemic state—can propagate independently and lead to less rational behavior, such as fear-driven isolation or the spread of misinformation that increases infection rates.

A primary challenge in integrated modeling is the lack of robust social, behavioral, and operational data, which forces modelers to rely on simplifying assumptions.[3] Furthermore, a review found that many existing modeling assumptions reflected a limited understanding of the interplay between social and behavioral factors.[11,14] Modelers often face the complex task of developing computationally tractable formulations that harness data and theory on complex social phenomena while maintaining parsimony. Several established and growing body of psychosocial research provides theoretical grounding for these modeling efforts. The Health Belief Model (HBM)[15], Theory of Planned Behavior (TPB)[16], and Protection Motivation Theory (PMT)[17] are among the most widely applied frameworks to explain the social, cognitive and emotional drivers of protective actions. These theories underscore the role of perceived susceptibility, perceived severity, self-efficacy, and social norms in shaping behavioral intention and compliance. Yet, significant heterogeneity persists in how behavior change is incorporated across models, with no standardized criteria to guide the choice of mechanisms, functional forms or theoretical grounding[10]. The motivation of our study is thus to suggest a more structured framework with clarified specific mechanisms for integrating behavioral science into epidemiological modeling.[18,19]

In this context, our review systematically examines post-2020 infectious disease models that incorporate behavioral adaptation mechanisms, focusing on how they incorporate human responses to an epidemic through several distinct mechanisms that modify transmission pathways or care-seeking behavior. We classify these models according to their underlying behavioral constructs, adaptation mechanisms, and impacts on model parameters and structure. Building on this classification, we propose a unifying taxonomy organized around behavioral drivers, social scale, and memory structure to highlight dominant modeling paradigms, evaluate their empirical grounding, and recommend pathways for more rigorous integration of behavioral science into epidemiological modeling. By mapping models to established psychosocial constructs, our approach offers a novel framework for guiding future behavior-adaptation model formulation, comparative evaluation, and data collection strategies.

## 2 Methods

**Search Strategy.** A systematic search was conducted on November 26, 2025, using the Scopus and PubMed databases to identify literature integrating infectious disease transmission with behavioral adaptation. (Table 1) The search strategy utilized a Boolean logic string combining three primary concept clusters: (1) infectious diseases (e.g., "epidemic," "pandemic," "outbreak"), (2) mathematical modeling frameworks (e.g., "compartmental model," "agent-based model," "SEIR"), and (3) behavioral dynamics (e.g., "behavior change," "risk perception," "adaptive decision," "compliance"). The search was limited to English-language articles published between 2020 and 2025. To ensure comprehensive coverage, the database search was supplemented by manually screening references from six identified key review articles.[11,19-21,22,23] Figure 1 summarizes the systematic search, screening, and inclusion process used to identify behavioral change modeling studies.

Table 1. Database Search Strategy and Results for Mathematical Models Incorporating Behavioral Dynamics

| Database | Specific Search Strategy (Title/Abstract Terms and Limits) | Results as of Nov 26 2025 |
|---|---|---|
| Scopus | TITLE-ABS-KEY ( ( "infectious disease*" OR "communicable disease*" OR epidemic* OR pandemic* OR outbreak* OR "emerging infectious*" ) AND ( "mathematical model*" OR "compartmental model*" OR "dynamic model*" OR "transmission model*" OR "differential equation*" OR "deterministic model*" OR SEIR OR SIR OR SEIS OR MSEIR ) AND ( "behavio* change" OR "behavio* adapt*" OR "behavio* response*" OR "risk perception" OR "awareness" OR "media effect*" OR "social distancing" OR "compliance" OR "adherence" OR "information dependent" OR "adaptive decision*" OR "coupled contagion" ) ) AND PUBYEAR > 2019 AND PUBYEAR < 2026 AND ( LIMIT-TO ( OA , "all" ) ) AND ( LIMIT-TO ( DOCTYPE , "ar" ) ) AND ( LIMIT-TO ( LANGUAGE , "English" ) ) AND ( LIMIT-TO ( SUBJAREA , "MEDI" ) OR LIMIT-TO ( SUBJAREA , "MATH" ) ) | 983 |
| Pubmed | (("infectious disease*"[Title/Abstract] OR "communicable disease*"[Title/Abstract] OR epidemic*[Title/Abstract] OR pandemic*[Title/Abstract] OR outbreak*[Title/Abstract] OR "emerging infectious*"[Title/Abstract]) AND ("mathematical model*"[Title/Abstract] OR "compartmental model*"[Title/Abstract] OR "dynamic model*"[Title/Abstract] OR "transmission model*"[Title/Abstract] OR "differential equation*"[Title/Abstract] OR "deterministic model*"[Title/Abstract] OR SEIR[Title/Abstract] OR SIR[Title/Abstract] OR SEIS[Title/Abstract] OR MSEIR[Title/Abstract]) AND ("behavio* change"[Title/Abstract] OR "behavio* adapt*"[Title/Abstract] OR "behavio* response*"[Title/Abstract] OR "risk perception"[Title/Abstract] OR "awareness"[Title/Abstract] OR "media effect*"[Title/Abstract] OR "social distancing"[Title/Abstract] OR "compliance"[Title/Abstract] OR "adherence"[Title/Abstract] OR "information dependent"[Title/Abstract] OR "adaptive decision*"[Title/Abstract] OR "coupled contagion"[Title/Abstract])) AND 2020:2025[dp] AND English[la] AND "journal article"[pt] | 779 |
| Review article | Bedson et al. 2021; Gozz et al 2025; Hamilton et al. 2024; Lee et al 2025; Proverbio et al 2025; and Li & Xiao, 2025 | 152 |
| Total without duplications | | 1,274 |

Figure 1. The search and inclusion process for behavior change modeling studies.

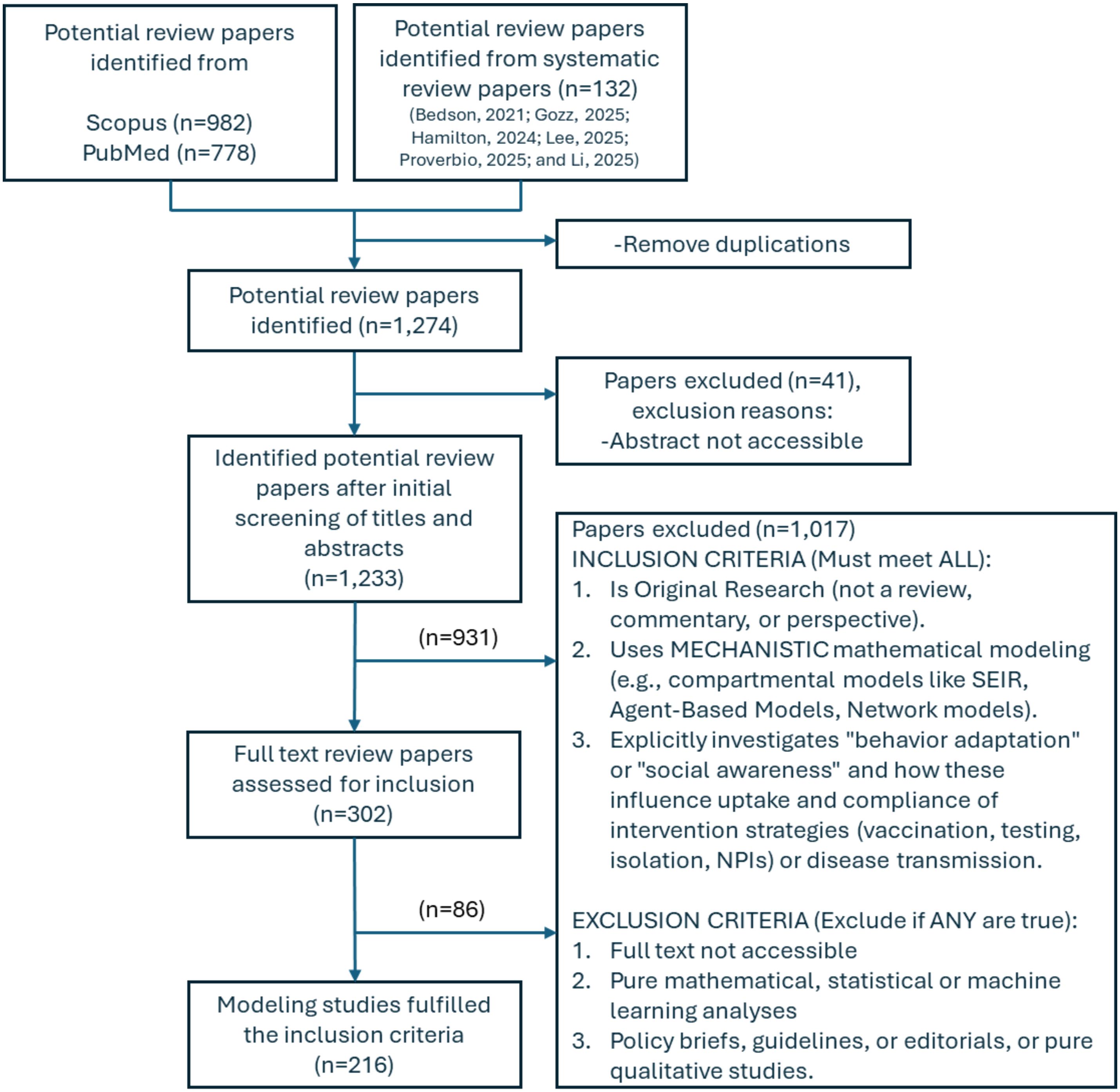

**Data Extraction.** Using a structured coding framework shown in Table S1, data were extracted from the included studies focused on both epidemiological and behavioral dimensions. Standardized fields included the specific country/region, the disease modeled, and the primary mathematical framework, such as game theory, agent-based models (ABMs), and compartmental ordinary differential equations (ODEs). A specialized extraction process was applied to analyze behavioral components, categorized by: (1) **Primary mathematical framework** used (e.g. model types: compartment ODE, ABM, network model, game theory); (2) **Behavior adaptation mechanism** (e.g., prevalence-dependent, peer imitation/social learning, media/awareness, exogeneous/policy driven, or game-theoretic); and (3) **Specific behavioral constructs** using a binary check for elements such as Personal Threat, Coping Appraisal, Structural Barriers, Social Norms, and Cues to Action. The key behavior constructions were informed by the major behavior change theories – Health Belief Model, Protection Motivation Theory, Theory of Planned Behavior, and Social Cognitive Theory. Additionally, studies were assessed for specific countries/regions modeled and the primary infectious disease being modeled.

Table 2 presents a structured taxonomy distinguishing behavioral mechanisms from behavioral constructs used in infectious disease models. Behavioral mechanisms characterize the *process by which behavior changes over time (i.e. how* behavior changes are generated), including endogenous responses that vary smoothly with epidemiological risk (incidence- or prevalence-driven)[24-27], responses triggered by external information such as media coverage or public health messaging[24-27], strategic decision-making framed as cost–benefit optimization in game-theoretic models, socially propagated behaviors arising through imitation or social learning[28-31], and exogenously imposed policy interventions that mandate behavioral change once thresholds are crossed.[32-34] Increasingly, these mechanisms are integrated within hybrid or multi-mechanism models that combine multiple pathways—such as adaptive feedback, social imitation, and policy triggers—to represent complex real-world behavioral dynamics. These models enable multi-scale behavioral responses and transitions between compliant and resistant states, giving rise to richer dynamics including fatigue, oscillations, and tipping points.[35,36,37] For example, high vaccination coverage may reduce perceived risk and prematurely relax social distancing, potentially triggering epidemic resurgence[38], while positive attitudes toward distancing can reinforce vaccine acceptance.[39 40] Such interactions illustrate how synergies—or tensions—between co-occurring behaviors can stabilize or destabilize epidemic trajectories, particularly when different behaviors respond to distinct information signals. Collectively, these mechanisms reflect a progression from fixed, exogenously specified behavior to dynamic, socially mediated formulations that more accurately capture human feedback shaping epidemic dynamics.[39,41]

On the other hand, behavioral constructs represent the cognitive, social, and contextual inputs that inform these behavioral processes. These constructs capture what individuals perceive, evaluate, and respond to when adapting their behavior, including perceived personal threat derived from biological indicators (e.g., incidence or mortality) or psychological states (e.g., fear or dread), coping appraisal of protective actions in terms of perceived efficacy, feasibility, and anticipated costs, and perceived barriers arising from economic hardship, social stigma, physical constraints, or behavioral fatigue over time. They also encompass social norms shaped by peer behavior, community expectations, and collective dynamics, as well as discrete cues to action such as policy announcements, media reports, or threshold-based alerts that precipitate behavioral change. Together, this classification separates the *inputs* to behavioral decision-making from the *mechanisms* that translate these inputs into action, clarifying both the drivers and implementation of behavioral adaptation in models and enabling systematic comparison of how behavior is conceptualized, triggered, and operationalized across the literature.

Table 2. Definition of behavior mechanisms and constructs used in behavior–disease models

| Category | | Definition & Context for Modeling |
|---|---|---|
| Mechanisms | Incidence/Prevalence Driven | -*Does behavior respond smoothly to perceived risk or information (disease burden)?* Behavior changes are triggered directly by the current number of cases, deaths, or perceived risk of infection in the population (e.g., people wear masks when cases rise, stop when cases fall). |
| | Media/Awareness-Driven | -*Does behavior respond smoothly to perceived risk or information (media/awareness on disease burden)?* Behavior is driven by external information campaigns, news reports, or public announcements (distinct from just reacting to case counts). |
| | Game-Theoretic | -*Is behavior a strategic response balancing costs and benefits?* Behavior is the result of a strategic payoff maximization (cost-benefit analysis) where individuals weigh the cost of infection vs. the cost of protection. |
| | Imitation/ Social Learning | -*Does behavior spread socially rather than rationally?* Behavior changes based on observing others or social pressure (e.g., "I will vaccinate if my neighbors do" or Evolutionary Game Theory dynamics). |
| | Exogenous/Policy | -*Is behavior triggered only after thresholds are crossed?* Behavior changes are forced or mandated by government interventions (e.g., lockdowns, school closures) rather than individual choice. |
| Construct | Personal Threat | The specific signals the agent monitors to assess danger, including biological data (prevalence, mortality counts) or psychological states (fear/dread). |
| | Coping Appraisal | The evaluation of the protective action itself, where agents weigh the strategy's utility (cost-benefit), its perceived efficacy (does it work?), and their capacity to perform it. |
| | Perceived Barriers | The frictions, costs, or obstacles that discourage protective behavior, spanning economic losses, social stigma, physical constraints, or psychological fatigue over time. |
| | Social Norms | The influence of peer behavior and community dynamics on the individual, driven by mechanisms such as imitation, herd compliance, or altruism. |
| | Cue to Action | Specific triggers—often external to immediate viral transmission—that prompt a change, such as government mandates (policy), media reports, or specific threshold events. |

**Results**

We summarized epidemiological and behavioral characteristics from 216 included studies, by publication year, geographic setting, disease context, modeling approach, adaptation mechanisms, and how behavioral processes influenced model structure and parameters. Figure 2 shows that publication volume peaked during the early COVID-19 pandemic (2020–2022) and declined thereafter. (Panel A) Studies were predominantly conducted in the United States (n= 87, 40%) and other high-income countries, with COVID-19 overwhelmingly dominating disease focus, while other infections were sparsely represented. (Panel B and C) Compartmental ODE models (n=174, 81%) constituted the majority of modeling approaches, whereas agent-based, network, and game-theoretic models were comparatively rare. (Panel D) Behavioral change was most commonly modeled through endogenous, incidence- or prevalence-dependent feedback mechanisms (n=103, 47%), followed by exogenous media or policy-driven responses (n=94, 44%); imitation and social learning mechanisms were infrequently incorporated. (Panel E) Behavioral adaptations most frequently targeted contact or transmission rates (n = 196, 91%), either as the primary focus alone or in combination with additional processes such as vaccination uptake or treatment dynamics; substantially fewer studies modeled vaccination uptake, adherence to non-pharmaceutical interventions, or treatment uptake as the primary behavioral target. (Panel F) We cite representative and methodologically illustrative studies in the main text. The complete list of all included studies (n = 216), along with key characteristics, is provided in Appendix Table S1.

Figure 2. Summary of included studies in the behavioral infectious disease modeling literature (N = 216)

A. Number of papers by year

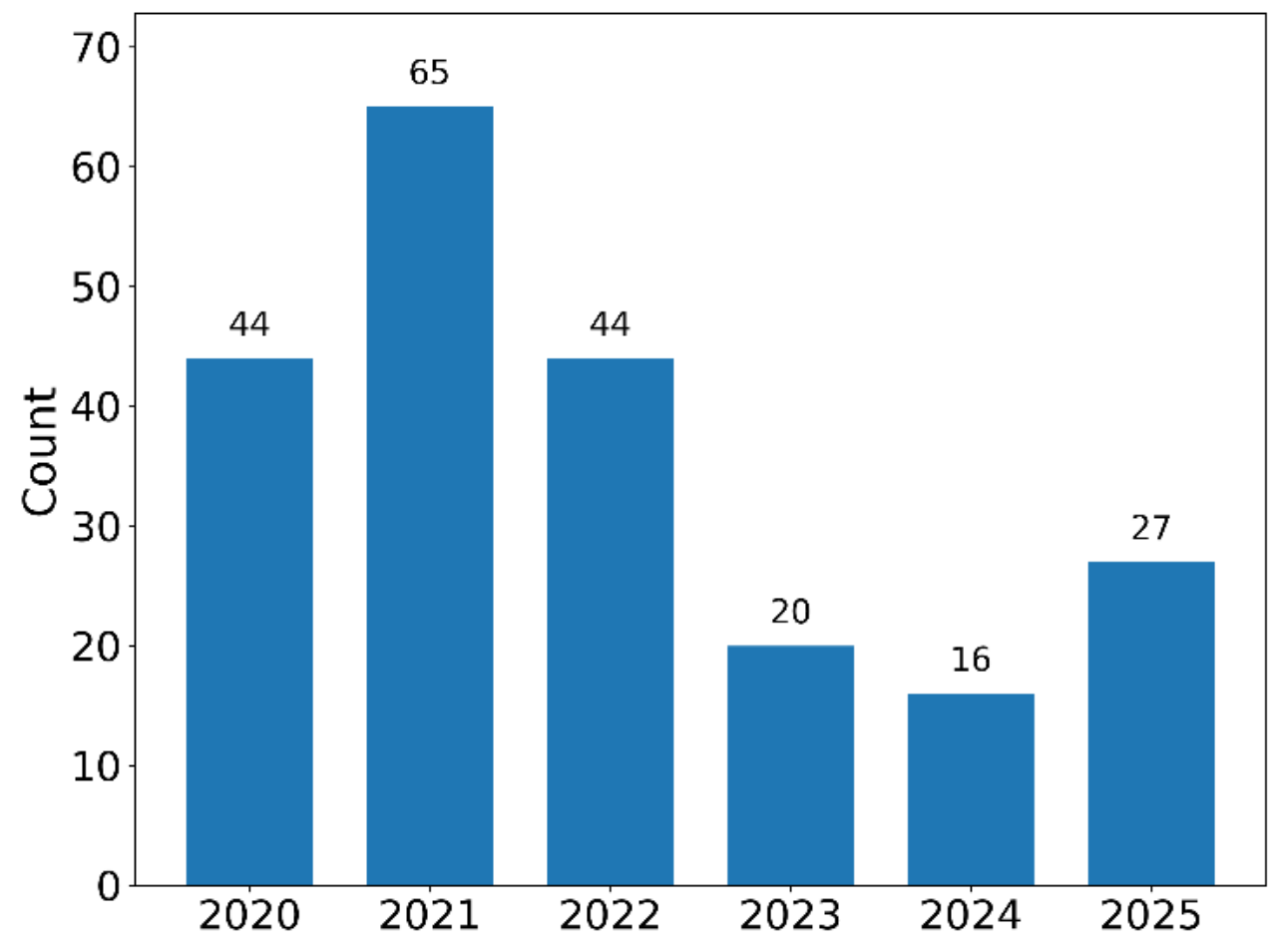

B. Number of papers by country

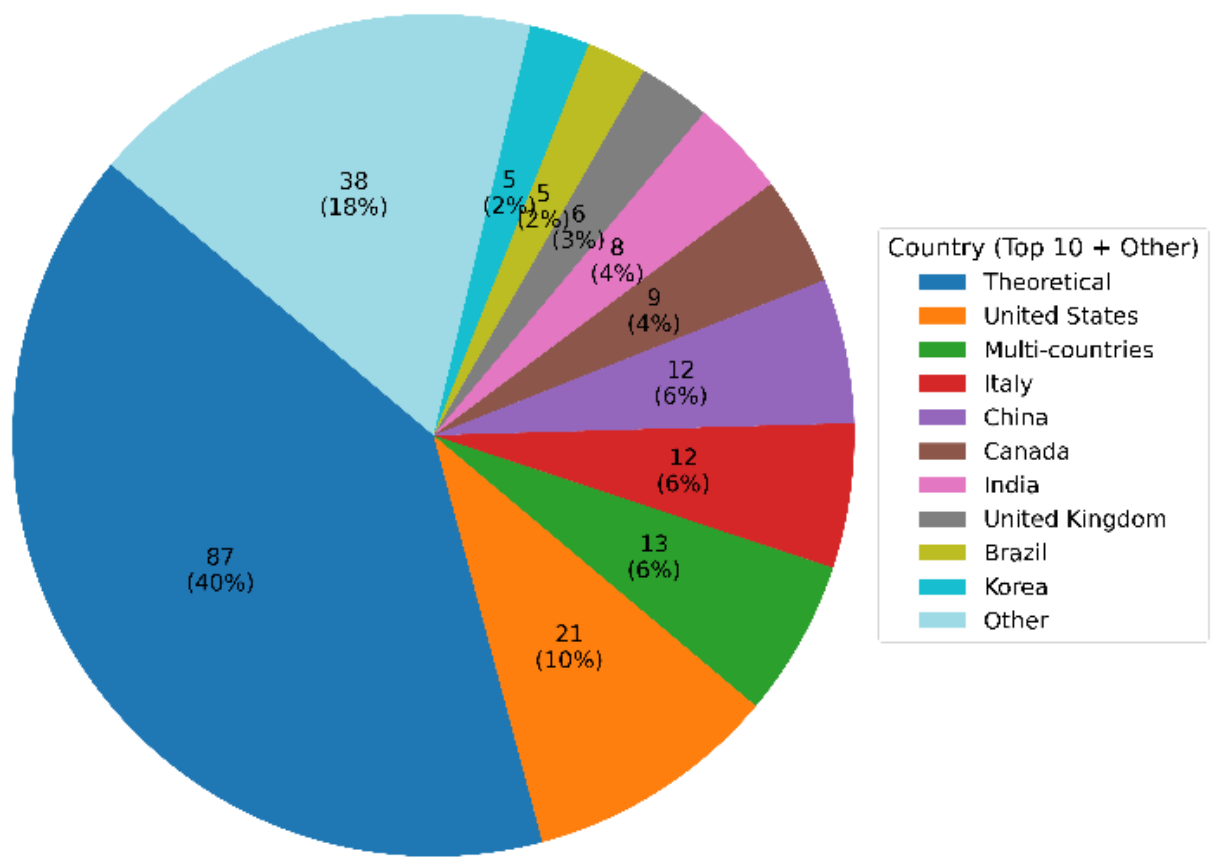

C. Number of papers by disease types

D. Number of papers by model types

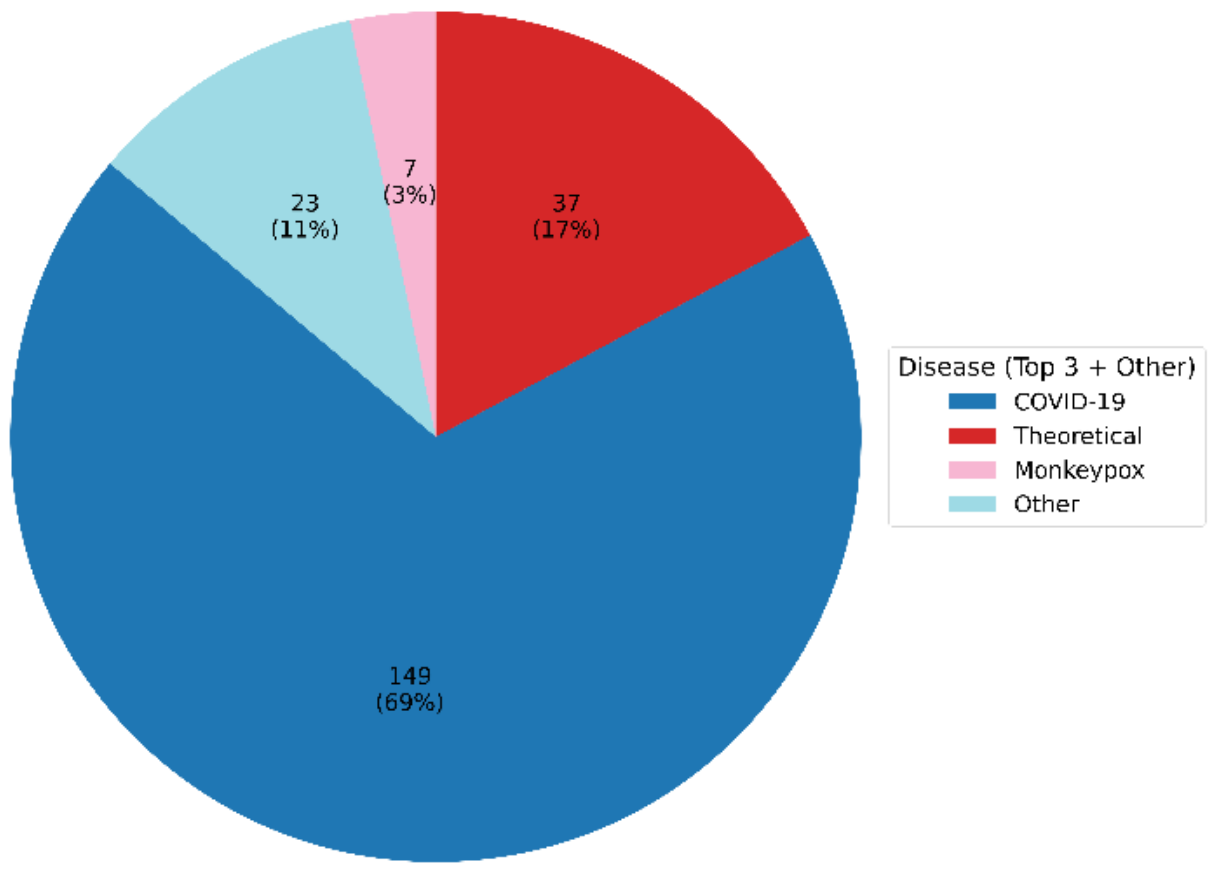

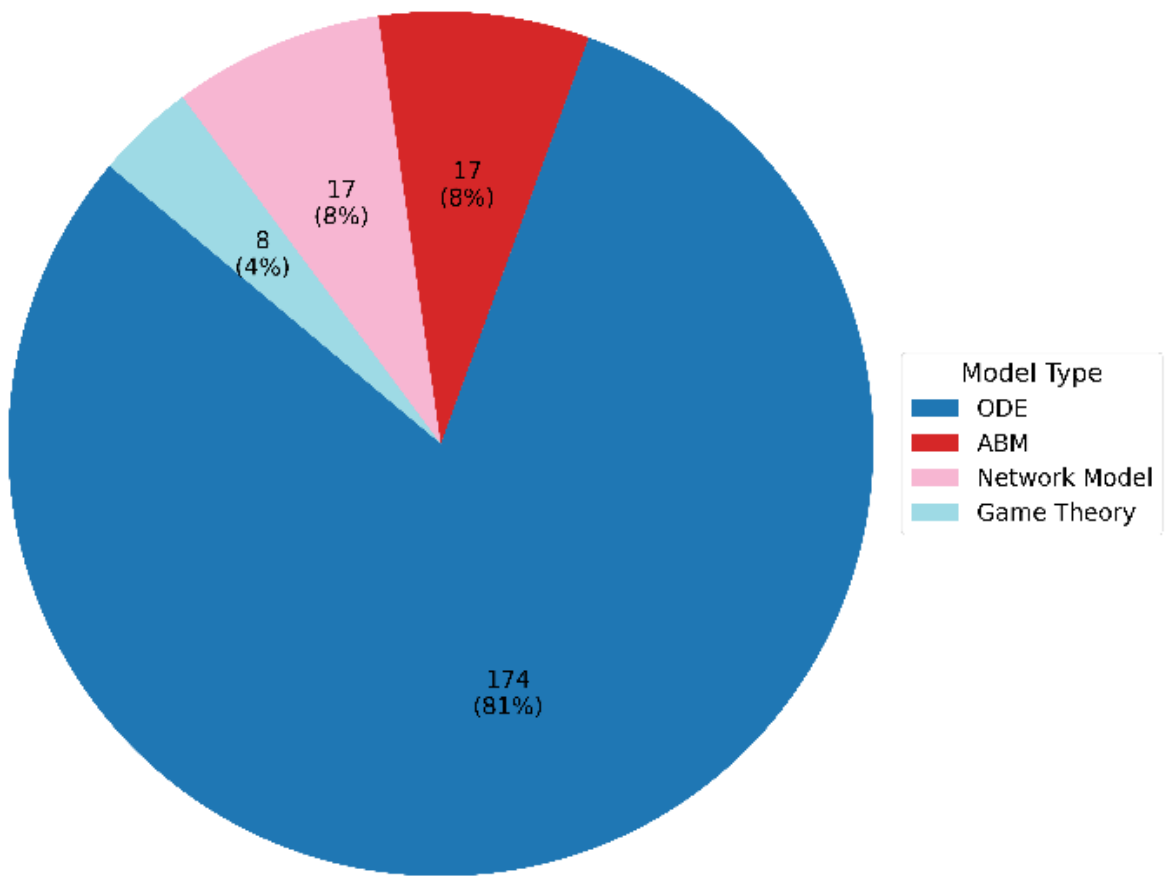

E. Number of papers by behavior mechanisms

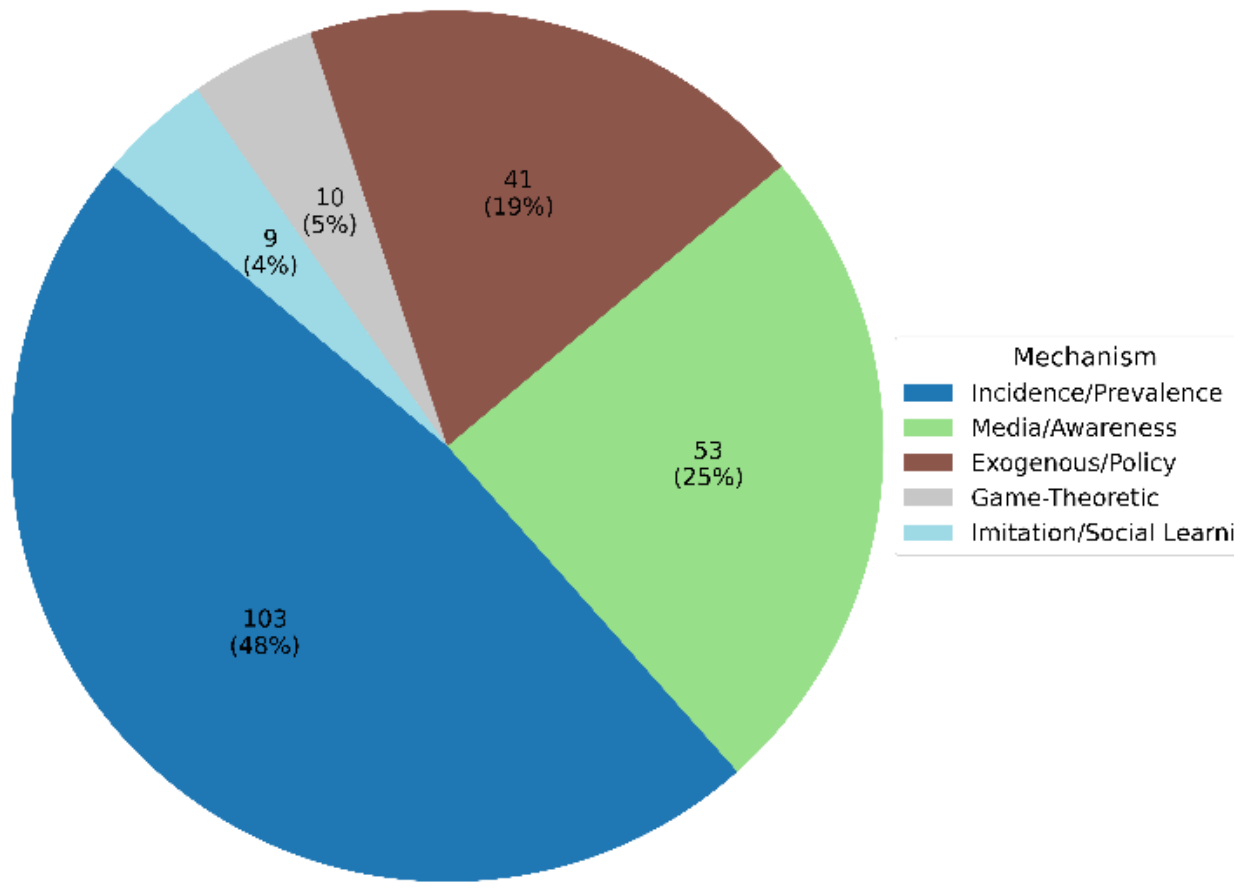

F. Number of papers by target variables

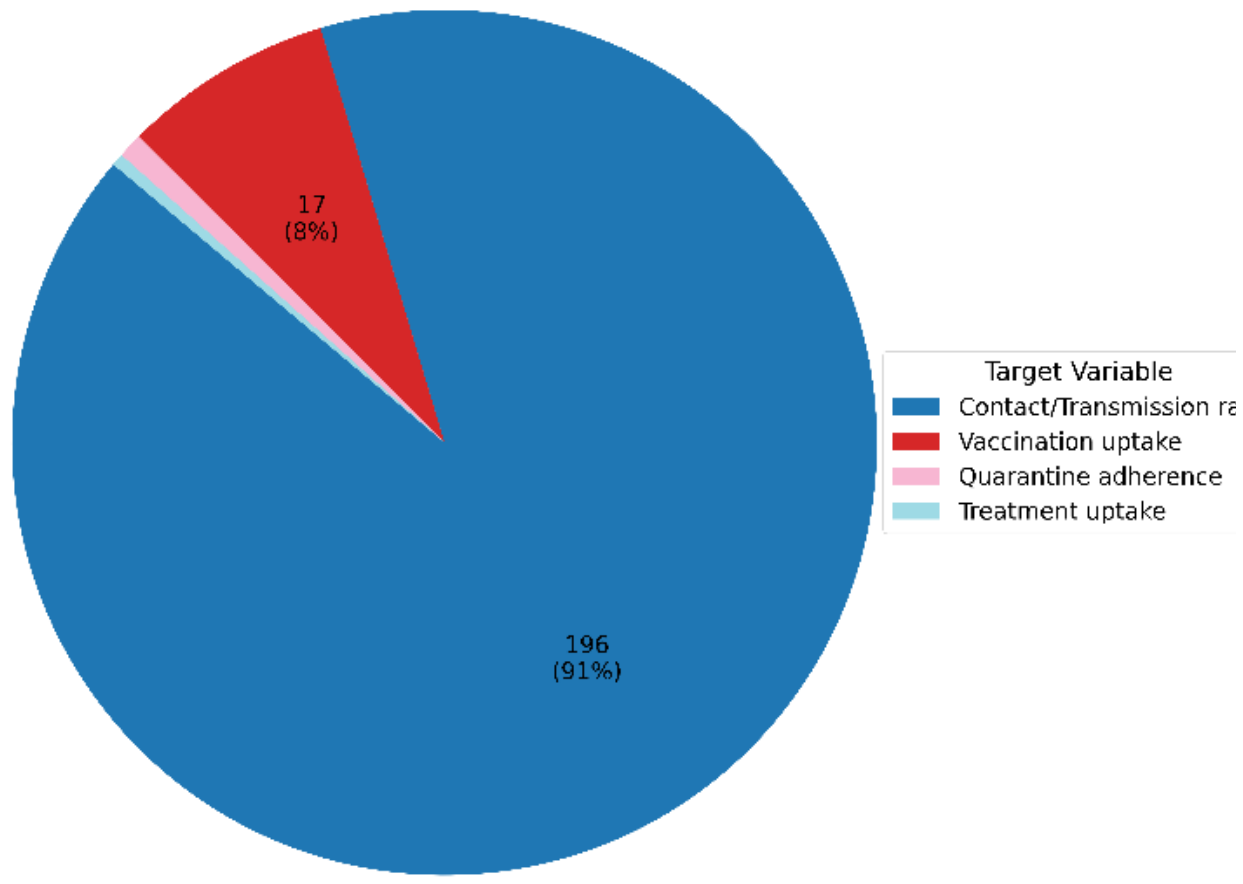

Note: Some studies incorporate multiple diseases or data sources, model types, behavioral mechanisms, and target variables. For example, certain papers analyze both influenza and COVID-19, or model behavior affecting both contact/transmission rates and vaccination uptake. For synthesis purposes, we identified a single primary model type, mechanism, and target variable for each study.

Figure 3 shows that most behavioral–epidemiological models rely on simplified, reactive mechanisms to represent human behavior. Cues to Action emerge as the most frequently modeled factor (N=159), predominantly as external triggers like government mandates[42] , policy announcements[43], media campaigns[44] , or social media signals[45]. Perceived personal threat is the most common endogenous driver (N=145), typically modeled as a decline in contact or transmission rates in response to rising cases, hospitalizations, or deaths—indicating that behavior is largely framed as fear-driven and reactive. In contrast, more complex constructs show substantial underrepresentation: Coping Appraisal (beliefs about intervention effectiveness, N=82) appears less than half as often, focusing on perceived efficacy of interventions like masking or vaccination, while Barriers (N=36) and Social Norms (N=25) are rarely included, indicating that structural constraints, compliance fatigue, and peer influence are largely overlooked in mechanistic transmission models.

These constructs are implemented in mechanistic but uneven ways. Cues to action are often implemented as step functions, time-dependent awareness variables, or exogenous shocks that abruptly alter transmission parameters.[46 47-50] Personal threat is formalized through epidemic-state-dependent functions that suppress transmission by eliciting public reactions to disease prevalence or severe outcomes.[51-53] Coping appraisal incorporates efficacy beliefs through parameters encoding intervention utility[45], benefit-cost trade-offs/payoff structures in game-theoretic frameworks[12,54], adherence strength tied to reduced disease severity[55], or responsiveness coefficients that govern the sustainability of protective behaviors once adopted.[56] Barriers and fatigue are occasionally included as costs, adherence decay, or socioeconomic constraints that limit behavioral change despite high awareness.[44,57,58] Social norms, though infrequently modeled, appear in network-based imitation dynamics, threshold adoption rules, or opinion amplification functions that propagate protective behaviors or concerns through peer interactions, marking a gradual shift toward viewing individuals as socially influenced agents rather than isolated decision-makers.[12,59,60]. Collectively, these patterns reveal a field that prioritizes policy- and fear-driven behavior while underrepresenting structural constraints and peer influence—suggesting an opportunity for next-generation models to better capture socially embedded, resource-constrained, and dynamically evolving human behavior.

Figure 3. Frequency of behavioral constructs represented across included studies (multiple constructs per study allowed)

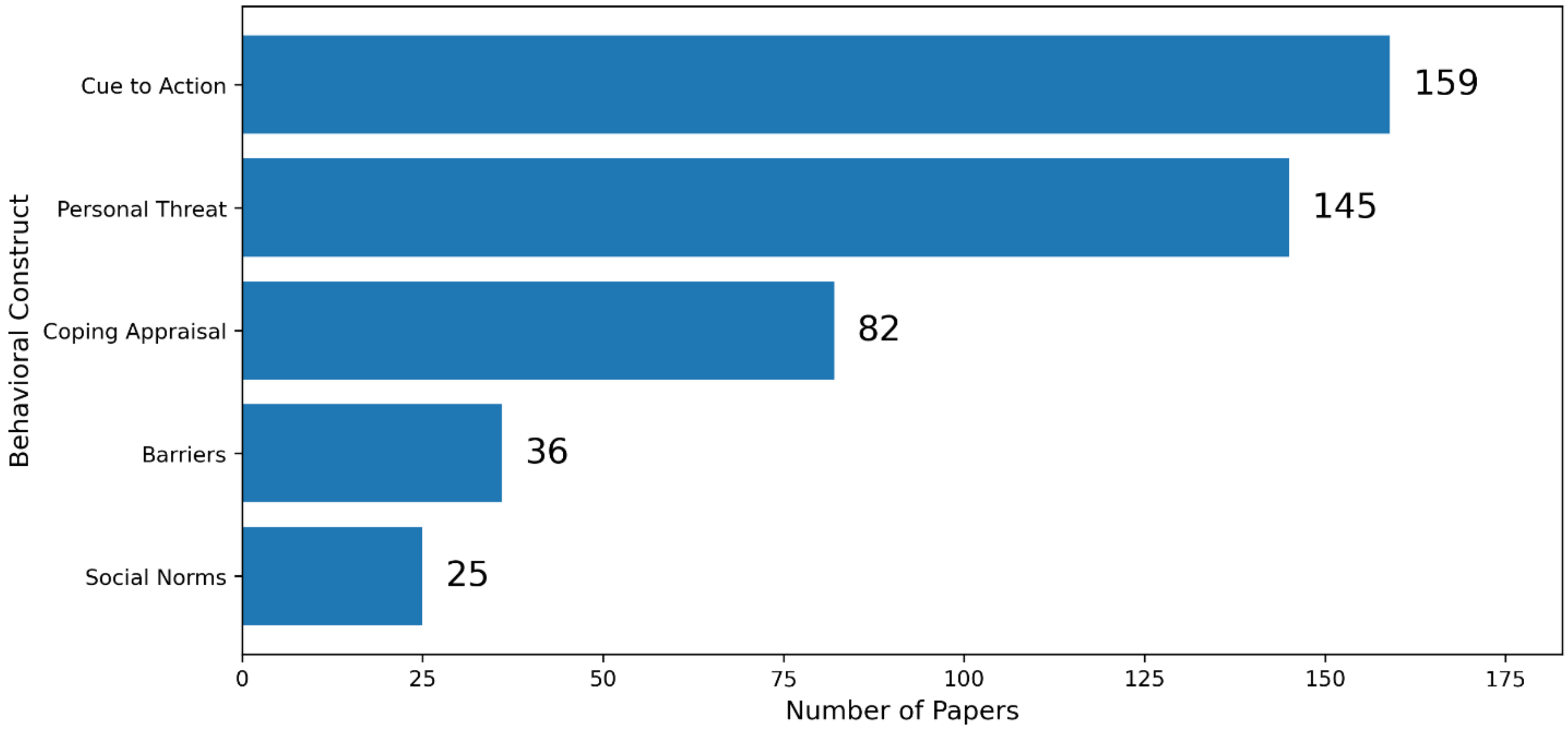

Table 3 mapped the reviewed studies onto a taxonomy defined by behavioral driver (endogenous vs exogenous), social scale (individual cognition vs social diffusion), and behavioral memory (short- vs long-memory) and links each category to its most common mathematical implementations. d'Onofrio, Manfredi, and collaborators[61,62,63] established the theoretical foundations of behavior–disease co-evolution by formally embedding awareness, information spread, and voluntary behavioral responses into transmission dynamics, and by analytically characterizing their rich dynamical properties, including oscillations and multiple equilibria. These works provide canonical examples of behavior–epidemic models and have continued to shape much of the contemporary literature, as summarized by Li and Xiao (2025)[23]. Endogenous, individual-cognition, short-memory models are by far the most common, reflecting their ease of implementation and modest data requirements. By embedding behavior directly into the force of infection (FOI) through prevalence-dependent or fear-based modifiers, these models provide a parsimonious way to represent rapid risk response and can be calibrated even when behavioral data are scarce. However, this parsimony comes at a conceptual cost: heterogeneous responses, persistence, and explicit behavioral decision-making are collapsed into a single functional form. As a result, distinct behavioral theories (e.g., fear, awareness, voluntary distancing) often map onto mathematically similar incidence modifiers, and behavioral parameters may not influence the basic reproduction number even when they strongly shape epidemic trajectories. Long-memory extensions at the individual level—implemented via delays, trend-sensitive perception, or switching—substantially enrich dynamics and can generate sustained oscillations and recurrent waves, but they introduce parameters that are difficult to identify empirically and are often weakly grounded in observed behavioral processes.

Models operating at the level of social diffusion offer greater behavioral expressiveness but at the cost of complexity and calibration challenges. These models modify FOI through strategy-specific mixing (e.g., vaccinated vs non-vaccinated contact rates) and can induce structural population splitting into behavioral sub-groups. Short-memory social-learning models, typically implemented through replicator or imitation dynamics, explicitly represent strategic adaptation and peer influence, allowing behavior to evolve endogenously and produce tipping points or multiple epidemic waves. Their strength lies in making behavioral feedback mechanisms transparent and interpretable in terms of incentives, yet they require strong assumptions about payoff structures and updating rules that are rarely empirically validated. Long-memory social-diffusion models, which encode persistent norms or resistance through behavioral subgroups, information stocks, or history-dependent learning, are conceptually well aligned with observed polarization and inequality in real epidemics, but remain comparatively rare and data-intensive. For exogenous drivers, piecewise or switching formulations are well suited for policy counterfactuals and scenario analysis but largely abstract away endogenous trust, fatigue, or adaptation, while media- and information-compartment models capture indirect and delayed effects of policy and communication but are highly sensitive to the chosen functional form linking information to behavior. Taken together, the table highlights that most behavioral diversity in the literature is realized through a limited set of mathematical primitives, underscoring the need for clearer justification of how specific behavioral assumptions are translated into equations—and for greater alignment between behavioral theory, data, and mathematical structure.

Table 3. Classification of behavior–epidemic models by behavioral driver, social scale, and memory structure, with corresponding mathematical formulations

| **Behavior Driver** | **Conceptual Definition** | **Mathematical Formulation & Examples (Adapted from Li & Xiao, 2025 )[23]** |
|---|---|---|
| | **Endogenous function (**state-dependent, feedback-driven) | |
| **Individual cognition** | Short-memory | |
| | • **Immediate risk response:** Behavior responds rapidly to current or recent epidemic signals (e.g., cases, deaths, hospitalizations).<br>• **Typical implementations:** prevalence directly dampens effective transmission in real time; fear or risk functions. $\lambda(\mathrm{S},\mathrm{I}) = \beta(\mathrm{I})\mathrm{SI}$ with $\beta$(I) decreasing in I<br>• **Strength:** analytically tractable; data-light; enables rapid calibration.<br>• **Limitation:** assumes homogeneous, proportional responses; weak representation of heterogeneity or persistence.<br>• *Representative studies:* Teslya et al. (2020)[64]; Tulchinsky et al. (2024)[52]; d'Onofrio, A. & Manfredi, P. (2022)[65]; d'Onofrio et al. (2022)[66] | **Modified incidence functions (saturation / nonlinear / exponential)**: Richer dynamics than bilinear incidence; behavior parameters may not enter $R_0$ in some forms, yet peaks/final size change; choice of functional form can qualitatively change dynamics (e.g., oscillations/bifurcations).<br>(1) Saturating contact rate: $\beta(I) = \beta_1 - \beta_2 \cdot I/(m + I)$<br>(2) Exponential incidence with behavioral suppression:<br>$\lambda = \beta IS/(1 + \alpha I^2)\ or\ \beta e^{-\alpha I} SI,\ \ \alpha$: behavioral sensitivity to prevalence<br>(3) Generalized nonlinear incidence: $\beta I^p S^q$ , $p, q$: nonlinearity exponents |
| | Long-memory | |
| | • **Fatigue, habituation, and trust erosion:** Behavior depends on cumulative epidemic experience and psychological adaptation over time.<br>• **Typical implementations:** behavior depends on lagged and/or trend signals; cumulative exposure or memory kernels; adherence decay functions.<br>• **Strength:** explains sustained oscillations, delayed resurgence, and compliance decay.<br>• **Limitation:** difficult to parameterize; behavioral states are often weakly constrained by data.<br>• *Representative studies:* Johnston & Pell (2020)[67]; Du et al. (2022)[68]; Espinoza et al. (2022)[69]; Kuwahara & Bauch (2024)[70] | **Delay differential equations (DDEs) or switching:** long-memory effects to internal decision processes, such as payoff-history learning or aspiration dynamics, where agents update behavior by comparing current outcomes to historical benchmarks.<br>(4) Delay response: $\lambda = \beta e^{-mI(t-\tau)} S(t) I(t)$ , τ: delay; m: sensitivity<br>(5) Trend-sensitive perception: $\lambda = \beta_0 e^{-M(t)} SI$<br>with $M(t) = \max\{0,\ \ p_1 I(t) + p_2 \frac{dI(t)}{dt}\}$, M: perceived signal; $p_1$,$p_2$: weights<br>(6) Switching: $\beta = \beta_0 e^{-\epsilon M_1(t)}$ with switching indicator ε=1_{S>S_c}, S_c: threshold<br><br>**Universal Differential Equation (UDE)/Neural ODE to learn hidden, history dependent mechanisms:** Behavioral effects on disease transmission are treated as unknown, data-driven functions and learned directly from epidemic and behavioral data using neural differential equations (NDEs / UDEs).<br>(7) Neural incidence / transmission function:<br>$\frac{dS}{dt} = -\mathrm{NN}(I, R)\ \frac{S}{N},\ \frac{dI}{dt} = \mathrm{NN}(I, R)\ \frac{S}{N} - \gamma I$<br>NN(·) is a neural network trained from data, representing the latent behavioral response. |

| **Social diffusion** | Short-memory | |
|---|---|---|
| | • **Rapid imitation and threshold cascades:** Behavioral change spreads endogenously through social interaction, information exchange, or adaptive contact patterns, rather than solely through individual risk evaluation.<br>• **Typical implementations**: Imitation/social learning, with payoff-based updating (replicator or imitation dynamics); Coupled disease–awareness diffusion, where spreading awareness feeds back to reduce contacts or transmission<br>• **Strength:** captures clustering, nonlinear adoption, and tipping points.<br>• **Limitation**: data- and computation-intensive; limited empirical calibration.<br>• Representative studies: Ye et al. (2020)[59]; Eksin et al. (2021)[71]; Huang et al. (2021)[72]; Mancastroppa et al. (2020)[73]; Juher et al. (2020)[74] | **Replicator & imitation dynamics (game-theoretic social learning)** can produce endogenous cycles/multiple waves and tipping-like adoption; equilibrium properties can differ sharply from incidence-modifier models.<br>(8) Replicator $\dot{x}_i = x_i[\pi_i(x) - \sum x_i\pi_i(x)]$ , $x_i$: strategy share; $\pi_i$: payoff<br>(9) General imitation $\dot{x}_i = x_i \sum_j [f_{ij}(x) - f_{ji}(x)] x_j$<br>$x_i$: the proportion of individuals adopting the strategy $i$, $f_{ij} = f_{ij}(f_i, f_j)$: the rate at which the $S_j$ switches to $S_i$<br>(10) Vaccination imitation: $\dot{x} = kx(1-x)[-r_v + r_i mI]$,<br>x: vaccinated fraction; $mI(t)$: probability of infection faced by unvaccinated individuals; $r_v$: cost of vaccination, $r_i$: costs of infection.<br>**Coupled disease–awareness diffusion** can generate endogenous oscillations, delayed responses, and multiple epidemic waves<br>(11) coupled diffusion $\dot{I} = \beta c(A)SI - \gamma I$, $\dot{A} = \kappa I(1-A) - \delta A$, $c(A) = C_0(1-\epsilon A)$ *or* $c(A) = \frac{C_0}{1+\epsilon A}$<br>$A$= awareness fraction; $\kappa$= awareness induction from infection; $\delta$= forgetting; $c(A)$reduces effective contacts/transmission. |
| | Long-memory | |
| | • **Persistent norms and resistance clusters:** Behavior becomes socially embedded, producing durable compliance patterns or polarized resistance.<br>• **Typical implementations**: durable social states and slow transitions encode norm persistence, polarization, and resistance by opinion dynamics; homophily-based networks; norm persistence models.<br>• **Strength:** captures polarization, inequality, and long-lasting behavioral stratification.<br>• **Limitation:** limited empirical validation; rarely integrated with policy design.<br>• Representative studies: Yedomonhan et al. (2023)[60]; Harris et al. (2023)[75]; Pei et al. (2024)[76]; Bentkowski & Gubiec (2025).[77] | **Persistent behavioral states coupled with information diffusion** can generate oscillatory dynamics or Hopf bifurcations even when behavior-related parameters do not alter the basic reproduction number $R_0$; highlights identifiability needs and data demands for long-memory social processes.<br>(12) Behavior transition by the total number of subgroups:<br>$\lambda \propto \alpha\frac{S^+ + I^+ + R^+}{N}$, $\alpha$ = per capita awareness transmission rate<br>(13) Awareness states: S (unaware) ⇄ $S^+$ (aware) with rates $\lambda SM$, $\lambda_0 S^+$,<br>Information memory M often follows $\frac{dM}{dt} = k\alpha I - \mu_0 M$, $\mu_0$: decay rate |
| | **Exogenous function (**externally imposed, independent of model states) | |
| **Individual cognition** | • **Mandates, incentives, and information shocks:** Behavior is imposed or triggered externally, largely independent of epidemic state.<br>• **Typical implementations:** parameters change abruptly based on thresholds or scheduled phases (policy on/off) by step functions; time-varying controls; optimal control formulations.<br>• **Strength:** policy-relevant; straightforward scenario evaluation.<br>• **Limitation:** neglects endogenous trust, fatigue, or behavioral adaptation.<br>• *Representative studies:* Zhang et al. (2020)[32]; Kennedy et al. (2020)[78]; Kraay et al. (2021)[33]; Baba & Bilgehan (2021)[79]. | **Piecewise / switching systems (policy shocks)**: outcomes can depend strongly on trigger placement; can create policy-driven cycling and new endemic states under repeated switching.<br>(14) Threshold-triggered policy regime$\beta(t) = \beta_0$ if $I < I_c$, $\beta_{lockdown}$ if $I \geq I_c$ , I: infected; $I_c$: policy threshold<br>(15) On–off behavioral suppression (Filippov-type formulation)<br>$\beta(t) = \beta^0 e^{(-\varepsilon(t)M(t))}$, switch indicator $\varepsilon(t) \in \{0,1\}$<br>$\varepsilon(t) = 0$ if $g(S,I,M) \leq 0$; $\varepsilon(t) = 1$ if $g(S,I,M) > 0$, M(t): policy/media intensity |
| **Social diffusion** | • **Media-driven norm shifts:** Policies or media reshape behavior indirectly through social diffusion and awareness.<br>• **Typical implementations**: information accumulates and decays; feeds back to reduce contacts/transmission by awareness diffusion; opinion dynamics; coupled information–disease models.<br>• **Strength:** captures indirect and delayed policy effects.<br>• **Limitation**: often simplified; norm formation rarely modeled explicitly.<br>• Representative studies: Buonomo & Della Marca (2020)[34]; Abueg et al. (2021)[80]; Zhao et al. (2021)[81]; Brand et al. (2023)[82]. | **Coupled media / awareness compartments (information stock M or social media proxy)**: infection→information→incidence feedback can generate oscillations/multiple outbreaks.<br>(16) Media compartment $\dot{M} = \eta I - \mu M; \beta_{eff} = \beta e^{(-\alpha M)}$, η: reporting; α: media impact<br>(17) SEIR +media $\lambda = e^{(-pM)} \beta SI/N; \dot{M} = \eta\delta E - \mu M$, δ: progression<br>(18) Joint suppression $\lambda = e^{(-\alpha_1 I - \alpha_2 M)} \beta SI$, $\alpha_1$, $\alpha_2$: sensitivities<br>(19) Rational response: $\lambda = \beta SI/(1 + \alpha_1 I + \alpha_2 M)$, $\alpha_1$,$\alpha_2$: response strengths<br>(20) Social media proxy: $\dot{T} = \mu_1 S + \mu_2 E + \mu_3 I - \tau T$, T: tweet volume<br>(21) Tweet-scaled incidence: $\lambda = e^{(-\alpha T)} \beta SI$, α: sensitivity |

Note: Endogenous function is a function whose value is determined by the internal state variables of the model and evolves through feedback within the system. Exogenous function is a function whose value is determined by external inputs or time-varying signals that are not governed by the model dynamics. Individual cognition: Behavioral change driven by an individual's own perception of epidemic risk, based on the current or past disease state, without explicit interaction or imitation of others' behavior. Social diffusion: behavioral change driven by exposure to others' behaviors or information, spreading through interpersonal contact, social influence, or population-level awareness.

## Discussion

This systematic review reveals a convergence in how behavioral adaptation has been incorporated into infectious disease transmission models over 2020-2025. Despite growing recognition that human behavior is shaped by cognitive, social, and structural processes, most models continue to rely on relatively simple, reactive formulations in which behavior responds proportionally to contemporaneous epidemic indicators such as case counts or deaths. Across 216 studies, behavioral change was overwhelmingly operationalized as prevalence- or incidence-dependent modulation of contact or transmission rates, with cues to action and perceived personal threat dominating the modeled behavioral constructs. In contrast, key dimensions emphasized in behavioral science—such as coping appraisal, structural barriers, social norms, and socially mediated learning—were sparsely represented and rarely embedded as dynamic, endogenous processes. This pattern mirrors findings from earlier conceptual reviews[14,18,83] but demonstrates that, even after the COVID-19 pandemic catalyzed unprecedented modeling activity, the field has largely favored analytically convenient and data-light behavioral representations over theoretically rich or socially grounded formulations. By classifying these diverse approaches into a unified taxonomy and framework of endogenous versus exogenous drivers, this review demonstrates that while current models excel at capturing *acute* "risk-response" responses, a critical methodological gap remains in quantifying the *sustained* and *socially complex* dynamics, such as pandemic fatigue, trust erosion, and the entrenchment of social norms that determine the long-term trajectory of epidemics. As a result, diverse behavioral theories are often collapsed into mathematically similar incidence modifiers, limiting the interpretability, comparability, and policy relevance of model-based insights.

In Table 3, we demonstrate how theoretical behavioral classifications can directly inform empirical model implementation, particularly for endogenous drivers that dominate both bodies of work. Our classification complements existing modeling reviews that are primarily organized around mathematical structure. For instance, Li and Xiao (2025)[84] categorize behavior–epidemic models by how behavioral feedback is embedded in transmission dynamics, including incidence modification, behavioral compartments, and evolutionary game formulations. Our framework, however, centers on the behavioral adaptation processes that motivate these formulations, organizing models by behavioral driver, social scale, and memory structure. Viewed together, these perspectives clarify that identical mathematical formulations may rest on fundamentally different behavioral assumptions, implying that model structure alone is insufficient for interpreting behavioral mechanisms. By explicitly linking behavioral theory to mathematical implementation, the framework enables clearer interpretation of model assumptions, more transparent comparison across studies, and more principled selection of behavioral mechanisms for policy-oriented epidemic modeling. Taken together, this taxonomy provides a unified and extensible language that supports theory-driven model design and cumulative evidence synthesis.

Advancing the field therefore requires treating behavior not merely as a reactive modifier of transmission, but as an endogenous, socially mediated, and temporally structured process that co-evolves with policy, information, and inequality. Applying the taxonomy across existing models reveals that the predominance of short-memory, individual-level formulations reflects modeling convention rather than empirical necessity. In doing so, the framework highlights behavioral processes that are routinely simplified or omitted—most notably socially mediated responses and long-memory effects—despite evidence of their influence on compliance persistence, polarization, and delayed epidemic resurgence. Importantly, the taxonomy provides a common basis for comparing models that differ in structure but share implicit behavioral assumptions, enabling systematic evaluation of how these assumptions shape parameterization, inference, and scope. For future studies, the framework offers practical guidance for aligning behavioral mechanisms with research objectives: short-memory individual responses may be adequate for near-term forecasting, whereas questions concerning intervention durability, equity, or behavioral fatigue require explicit representation of social diffusion or memory-dependent dynamics. In this way, the taxonomy shifts the focus from whether to include behavior toward how specific behavioral processes should be incorporated to enhance explanatory clarity and policy relevance.

**Behavior as a driver of multiple waves and oscillations.** A primary finding across these models is that human behavior is a critical, endogenous driver of epidemic dynamics, particularly in the formation of multiple waves and oscillations. Models that incorporate behavioral feedback, especially those accounting for time delays and psychological fatigue, naturally replicate the recurring waves seen in real-world outbreaks[85,86]. This demonstrates that secondary waves are not random events but are often a direct and predictable consequence of the population's collective actions.[51] This phenomenon is driven by key mechanisms, including system delays—such as lags in public response or the flow of information—which create "catch-up" cycles where behavior lags behind the infection, inducing periodic oscillations.[87-89] This is compounded by fatigue and frustration, where waning adherence to restrictions or the simple decay of awareness leads to a relaxation of protective behaviors that directly trigger a resurgence.[90] Finally, these cycles are reinforced by changes in risk perception: when interventions reduce case numbers, people perceive less risk, relax protective behaviors, and new waves can emerge. Whether this happens depends on whether the public anticipates future risk ("forward-looking") or responds only after cases rise ("myopic").[12,91,92]

**Behavior as a choice of payoffs or social learning.** Models that treat behavior as a payoff-driven choice or a social learning process highlight how underlying network structures and information flows act as critical levers in epidemic dynamics.[93,94] Imitation often produces nonlinear—and sometimes counterintuitive—responses: as a protective behavior becomes widespread, infection risk declines, making defection more appealing and potentially generating endogenous cycles of high and low compliance or multiple epidemic waves.[95] These dynamics can also be highly sensitive to the payoff structure; even small shifts in perceived costs and benefits—such as a modest subsidy, a slight increase in perceived vaccine risk, or incorporating altruistic concern for others—can move the system between high- and low-uptake equilibria or create tipping points. Imitation-based frameworks are most informative when social influence is a central driver of behavior (e.g., vaccine hesitancy, masking norms, sexual risk-taking in dense networks), when heterogeneous or clustered patterns matter more than population averages, and when policy levers operate through incentives or norms rather than mandates.[96] Conceptually, these models encourage moving beyond the question of how much behavior changes with prevalence to asking who imitates whom, over which network, under what incentives, and how those social learning rules feed back to shape epidemic trajectories.

**The power of information and media.** Information—whether accurate or misleading—emerges as one of the most powerful drivers of behavioral change in epidemic dynamics, functioning as a primary *cue to action* that shapes individual and collective responses.[97] Across studies, media coverage, public health campaigns, and social media platforms were found to be critical in modulating awareness, influencing perception of risk, and ultimately determining the trajectory of disease spread.[98] Timely, high-coverage, and sustained information dissemination through advertisements, educational programs, or public service announcements substantially reduced both epidemic peaks and overall case burden by promoting protective behaviors.[45,64,84,87,99] Several modeling studies suggest that the effects of communication and awareness on epidemic control are nonlinear. While insufficient information fails to mobilize protective behavior, increasing exposure often exhibits diminishing returns and, in socially heterogeneous settings, may even amplify confusion, polarization, or disengagement—ultimately weakening collective response.[75,100] Moreover, digital platforms such as Twitter and Google Trends provide early warning signals of declining vaccine sentiment or emerging resistance, often detectable years before outbreaks materialize.[101,102] Beyond cognitive awareness, emotional contagion—fear and panic spreading through social networks—acts as an endogenous feedback mechanism, prompting spontaneous self-protective behaviors that can significantly flatten epidemic curves.[,67,85,103] Together, these findings highlight that the information environment is not merely a backdrop to epidemics but a central, dynamic force that shapes behavioral trajectories, risk perception, and epidemic outcomes.

**The critical importance of timing and combined interventions.** The models clearly show that an intervention's effectiveness is critically defined by when it is deployed and what it is paired with. Timing is a dominant factor, with early interventions—whether education, lockdowns, or awareness campaigns—proving vastly more effective than those implemented late.[104] For example, models showed that early implementation can substantially reduce cumulative infections, whereas introducing the same campaign later in an outbreak has much smaller effects.[73,105] Beyond timing, the findings highlight that interventions are most effective when combined in a synergistic, multi-layered approach.[106,107] The models found greater success when pairing public awareness with sanitation, combining vaccination with self-protection training, or supporting contact tracing with robust physical distancing.[108,109] This synergistic effect is particularly evident in the interplay between top-down policy and bottom-up public response, where government-mandated actions and voluntary individual action are mutually reinforcing and essential for controlling an outbreak.

Importantly, the insights generated by behavior-adaptive models extend beyond infectious disease prediction. These models have increasing relevance for addressing today's most complex public health challenges, including vaccine hesitancy, declining institutional trust, misinformation, chronic disease management, climate-related risk responses, and the design of equitable interventions. Because they embed behavior within dynamic systems, such models provide a unique lens for identifying tipping points in collective action, anticipating unintended behavioral responses to policy, and designing interventions that harness social influence or reinforce adaptive behavior. They also offer a laboratory for testing foundational social and behavioral theories under dynamic, real-world feedback conditions—something that static behavioral frameworks cannot accomplish. By integrating constructs from the Health Belief Model, Protection Motivation Theory, Theory of Planned Behavior, and Social Cognitive Theory with epidemic processes, behavior-adaptive models can reveal how risk perception, self-efficacy, social norms, and fear-based responses evolve over time and interact with epidemic dynamics.

**Limitations**. This review has several limitations. Although our search focused on behavior-relevant keywords, the absence of standardized terminology may have led to the exclusion of relevant studies that employ different definitions.[110] Such conceptual challenge highlighted by our review is that, in a broad sense, all epidemic models contain behavior, because human contact, mobility, and care-seeking patterns are inherently behavioral. This raises the question of where to draw the boundary between “behavior models” and conventional epidemic models that simply assume modified parameters. Our review suggests that the distinction is not binary but a continuum of behavioral integration**,** ranging from models with fixed behavioral assumptions to those where behavior-related variables evolve endogenously through dynamic feedback loops, social imitation, or policy-driven cues. For the purposes of synthesis, we defined behavior-adaptive models as those in which behavioral indicators dynamically modify epidemic processes within the system over time, rather than being treated as static modifiers. Even under this definition, ambiguity persists—particularly for models where behavior changes are triggered by policy interventions rather than by autonomous individual or social responses. This ambiguity signals a deeper need for conceptual clarity and standardized definitions in future research. Finally, although our synthesis spans a wide temporal range, the literature on behavior–disease modeling is rapidly evolving, and recent studies employing hybrid frameworks, real-time behavioral data streams, and machine-learning calibration methods[111] may not yet be fully represented. Despite the limitations, our study offers practical guidance for translating psychological drivers into dynamic uptake functions, adherence states, and policy-triggered step changes that can shift epidemic peaks and reduce incidence.

**Recommendation.** Human behavior is inherently complex, shaped by cognitive, emotional, social, and structural factors that interact in nonlinear ways. Because no single model can feasibly capture this full spectrum, the goal of behavior–disease modeling should be purposeful representation, not exhaustive inclusion[19]. Modelers must prioritize the behavioral mechanisms that most directly influence epidemic outcomes or policy decisions. Simplified adaptive functions, such as prevalence-dependent contact reduction or awareness decay, may suffice for large-scale forecasting, while more detailed frameworks—like agent-based or network imitation models[76,77]—are warranted when exploring heterogeneity, clustering, or social contagion of behaviors[112,113]. Clarifying the appropriate level of abstraction is essential to balance tractability, interpretability, and empirical relevance.[51] Overly complex models risk becoming computationally opaque or uncalibratable, while oversimplified ones may neglect critical feedback between perception, policy, and transmission[10,114]. By explicitly defining what behavioral components are included and why, modelers can align complexity with the precision required by the question at hand. This focus allows analytical and computational resources to be directed toward mechanisms that meaningfully alter outcomes, ensuring the model’s insights remain actionable for public health decision-making.

Future behavior–epidemic modeling studies should move beyond predominantly mechanism-driven demonstrations toward systematic data integration and rigorous calibration and validation**.** A review by Bedson et al. of 178 articles[19] found that fewer than 20% incorporated real behavioral data for parameterization, relying instead on theoretical assumptions. Similarly, Hamilton et al.[11] found that while calibration was more common in recent COVID-19 models, true behavioral validation remained rare, with many studies using epidemiological proxies rather than direct behavioral measurements. Recent COVID-19 models, however, show that data-informed behavioral modeling is feasible[21],[20] nearly 70% fitted their models to observed data, and about a quarter compared forecasts to real-world trends.[11] Building on this progress, we recommend that studies clearly distinguish which behavioral components are empirically informed versus assumed, and prioritize calibration of behavioral parameters using available data sources such as epidemiological time series, mobility data, and behavioral surveys. Equally important, validation practices should be strengthened by routinely comparing behavior-integrated models with non-behavioral baselines fitted to the same datasets, rather than relying primarily on sensitivity analyses or qualitative comparisons. Researchers should rigorously assess the "value-add" of behavioral complexity by fitting both the behavior-integrated model and a baseline non-behavioral version to the same observed dataset. This comparative approach is essential to distinguish whether improved fits are due to genuine behavioral insights or simply increased model flexibility.

Ultimately, aligning model complexity with policy relevance requires asking which behaviors (contact/vaccination; voluntary/policy driven)—and at what scales (individual/group/societal)—change epidemic trajectories enough to matter. For example, individual risk perception may drive vaccination decisions, while social norms and peer influence shape compliance with masking or distancing[115]. Purposefully selecting these levers enhances the model's explanatory and practical value, helping policymakers identify where communication, resource allocation, or structural interventions will be most effective. This pragmatic, theory-informed approach ensures that behavioral realism strengthens rather than obscures understanding, allowing models to remain both scientifically credible and directly useful in guiding public health strategies. Advancing this field requires constructive interdisciplinary dialogue between modelers, psychologists, economists, and communication scientists[115]. Behavioral scientists can translate constructs like self-efficacy, norms, and trust into measurable variables, while modelers can clarify how these constructs are operationalized or estimated from data. This collaboration transforms calibration into a process of scientific negotiation—testing assumptions, refining constructs, and aligning theoretical and empirical representations[19].

**Conclusion**

Human behavior plays a central and dynamic role in shaping infectious disease transmission, yet integrating these processes into epidemiological models remains one of the most conceptually challenging and methodologically heterogeneous areas in disease modeling. In many socio-epidemiological models, behavioral parameters are linked to real-world data in one of two ways: (i) by fixing epidemiological parameters and estimating behavioral ones directly from epidemic time series—a method often limited by weak identifiability when relying solely on epidemic curves; or (ii) by calibrating models using observable behavioral proxies, such as mobility data, Google Trends, social media indicators, or policy indices, occasionally complemented by survey data when available. While these approaches offer a bridge to reality, they often serve as rough proxies for complex cognitive processes, underscoring the need for more rigorous, data-driven parameterization. Our review advances the field by shifting the focus from whether behavior matters to how it should be represented in ways that are theoretically coherent, empirically grounded, and decision-relevant. The taxonomy developed here provides a structured lens for navigating this design space, making explicit the trade-offs between parsimony and behavioral realism, and clarifying how different assumptions about cognition, social influence, and memory are translated into mathematical form. By distinguishing behavioral drivers, social scale, and temporal structure, the framework enables more transparent interpretation of existing models and more intentional alignment between research questions, data availability, and model complexity. Looking ahead, progress in this field will depend on closer integration between behavioral theory, emerging behavioral data streams, and model validation practices, as well as sustained interdisciplinary collaboration. Advancing behavior–epidemic modeling in this direction will be essential not only for improving outbreak forecasting, but also for designing interventions that are robust to fatigue, heterogeneity, and the social dynamics that ultimately determine public health outcomes.

**Data Availability.** All data produced in the present work are contained in the manuscript and supplementary material.

**Funding.** This research was supported by a New Faculty Startup Fund from UConn Health (grant number 208069-10100-72525-10). The funders had no role in study design, data collection and analysis, decision to publish, or preparation of the manuscript.

-

## Appendix

Table S1. Summary of Included Behavioral–Epidemic Modeling Studies and Coded Psychosocial Constructs

| ID/Reference number | First Author | Year | Disease | Mode_Type | Behavior_Mechanism | Target Variable | Psychosocial constructs | | | | | Country |
|---|---|---|---|---|---|---|---|---|---|---|---|---|
| | | | | | | | Personal_Threat | Coping_Appraisal | Barriers | Social_Norms | Cue_to_Action | |
| 1 | Milne AE | 2020 | Citrus Huanglongbing | ABM | Imitation/Social Learning | Contact/Transmission rate | Yes (Growers' perceived risk of orchard infection explicitly drives decisions) | Yes (perceived effectiveness of area-wide control determines willingness to participate) | No | Yes (opinions evolve through interactions with other growers and advisors via opinion dynamics) | Yes (extension-agent information pulses and direct observation of infection trigger opinion updates) | United States |
| 2 | Silva, P.C.L. | 2020 | COVID-19 | ABM | Exogenous/Policy | Contact/Transmission rate | No | No | Yes (Homeless agents cannot execute the 'Go home' action to isolate) | No | No | Brazil |
| 3 | Zhang, K. | 2020 | COVID-19 | ABM | Exogenous/Policy | Contact/Transmission rate | No | No | No | No | Yes (mask-wearing compliance and shelter-in-place policies) | United States |
| 4 | Glaubitz A | 2020 | COVID-19 | Game Theory | Incidence/Prevalence | Contact/Transmission rate | Yes (individuals decide on social distancing based on the risk of infection.) | Yes (individuals weigh the cost of social distancing against the risk of infection.) | No | Yes (social learning/imitation process is mentioned.) | No | Theoretical |
| 5 | Braun B | 2020 | COVID-19 | Network Model | Incidence/Prevalence | Contact/Transmission rate | No | No | No | No | Yes (self-isolation in response to symptomatic contacts) | Theoretical |
| 6 | Karaivanov A. | 2020 | COVID-19 | Network Model | Incidence/Prevalence | Contact/Transmission rate | Yes (based on observed infections or deaths among the agent contacts or in the population at large) | Yes (Agents reduce contacts if a social contact tests positive, representing a 'fear-driven' response to immediate risk) | No | No | Yes (Behavioral responses are triggered when aggregate new cases exceed a specific threshold over the preceding 20 days) | Theoretical |
| 7 | Komarova N.L. | 2020 | COVID-19 | Network Model | Incidence/Prevalence | Contact/Transmission rate | Yes (risk perception and fear of infection) | No | No | No | Yes (mandated social distancing) | United States |
| 8 | Ye, Y. | 2020 | COVID-19 | Network Model | Media/Awareness | Contact/Transmission rate | Yes (risk perception differences among individuals) | No | No | Yes (social influence on behavior change) | Yes (media and opinion leaders) | Theoretical |
| 9 | Cardoso EHS | 2020 | COVID-19 | ODE | Exogenous/Policy | Contact/Transmission rate | No | No | Yes (High number of people per bedroom and lack of water supply) | No | No | Brazil |
| 10 | Carcione, J.M. | 2020 | COVID-19 | ODE | Exogenous/Policy | Contact/Transmission rate | No | No | No | No | Yes (government-imposed lockdown and isolation measures implemented at specific times) | Malaysia |
| 11 | Ejigu B.A. | 2020 | COVID-19 | ODE | Exogenous/ Policy | Contact/Transmission rate | No | No | Yes (The model accounts for barriers such as lack of access to sanitation materials and different living conditions in rural vs urban areas affecting adherence) | Yes (Physical distancing is noted as a custom in rural areas, implying a social norm) | No | Ethiopia |
| 12 | Gupta, S.D. | 2020 | COVID-19 | ODE | Exogenous/ Policy | Contact/Transmission rate | No | No | No | No | Yes (social distancing and lockdown are implemented as public health interventions and explicitly parameterized in the model) | India |
| 13 | Yang T | 2020 | COVID-19 | ODE | Exogenous/ Policy | Contact/Transmission rate | No | No | No | No | Yes (governmental interventions such as travel bans, social distancing, school and workplace closure) | Theoretical |
| 14 | Makanda, G. | 2020 | COVID-19 | ODE | Exogenous/ Policy | Contact/Transmission rate | No | No | No | No | Yes (quarantine and social distancing policies) | Theoretical |
| 15 | Bozkurt F | 2020 | COVID-19 | ODE | Exogenous/ Policy | Contact/Transmission rate | No | No | No | No | Yes (police and health organizations) | Theoretical |
| 16 | Kouidere, A. | 2020 | COVID-19 | ODE | Exogenous/ Policy | Contact/Transmission rate | No | No | No | No | Yes (awareness campaigns, quarantine measures) | Theoretical |
| 17 | Khan ZS | 2020 | COVID-19 | ODE | Exogenous/ Policy | Contact/Transmission rate | No | No | No | No | Yes (stay-at-home orders, lockdowns) | United States |
| 18 | Althouse BM | 2020 | COVID-19 | ODE | Exogenous/ Policy | Contact/Transmission rate | No | No | Yes (travel restrictions and venue closures) | No | Yes (government policies and media announcements) | United States |
| 19 | Kennedy DM | 2020 | COVID-19 | ODE | Exogenous/ Policy | Contact/Transmission rate | Yes (engaging in personal protection measures like wearing face masks) | Yes (importance of personal protection measures) | No | No | Yes (policy mandates and recommendations by the CDC) | United States |
| 20 | Currie DJ | 2020 | COVID-19 | ODE | Exogenous/ Policy | Contact/Transmission rate | No | No | Yes (privacy concerns affecting app uptake) | No | Yes (government-initiated contact tracing app, testing policies, and social distancing mandates are imposed and parameterized as external interventions) | Australia |
| 21 | Acuna-ZegarraMA | 2020 | COVID-19 | ODE | Exogenous/ Policy | Contact/Transmission rate | No | Yes – explicit trade-off between compliance duration, abandonment (leak rate ?), and effectiveness of | Yes (partial adherence and abandonment of sanitary measures are explicitly modeled via population split and leak rate) | No | Yes (implementation of Sanitary Emergency Measures (SEM) and suspension of non-essential activities trigger behavior change | Mexico |
| | | | | | | | | contact rate reduction under sanitary measure | | | at a specified time point) | |
| 22 | Pedro SA | 2020 | COVID-19 | ODE | Imitation/Social Learning | Contact/Transmission rate | Yes (perceived risk of SARS-CoV-2 infection) | No | No | Yes (social learning dynamics) | No | Theoretical |
| 23 | Ferchiou, A. | 2020 | COVID-19 | ODE | Incidence/Prevalence | Contact/Transmission rate | Yes (people changed behavior due to awareness of health risks) | No | No | No | Yes (media raised awareness) | France |
| 24 | Gatto M | 2020 | COVID-19 | ODE | Incidence/Prevalence | Contact/Transmission rate | Yes (increased awareness after first cases) | No | No | No | Yes (government restrictions and media announcements) | Italy |
| 25 | Kim, S. | 2020 | COVID-19 | ODE | Incidence/Prevalence | Contact/Transmission rate | Yes (People changed behaviors such as avoiding close contacts or enhancing personal hygiene) | No | No | No | Yes (Government announcements and school closure policies) | Korea |
| 26 | Johnston MD | 2020 | COVID-19 | ODE | Incidence/Prevalence | Contact/Transmission rate | Yes (perceived fear of infection) | No | No | No | No | Multi-countries |
| 27 | Buckman, S.R. | 2020 | COVID-19 | ODE | Incidence/Prevalence | Contact/Transmission rate | Yes (Rising number of infections triggers a response) | No | No | No | Yes (Linked to mandatory and voluntary containment measures) | Multi-countries |
| 28 | Kim S | 2020 | COVID-19 | ODE | Incidence/Prevalence | Contact/Transmission rate | Yes (Behavior changes in response to increasing cases) | No | No | No | Yes (Government announcements and public campaigns) | Korea |
| 29 | Nyabadza F | 2020 | COVID-19 | ODE | Incidence/Prevalence | Contact/Transmission rate | No | No | Yes (The model assumes perfect social distancing is not attainable due to settlement patterns and high population density in informal settlements) | No | Yes (The model triggers the change in contact rate based on the specific date of the national lockdown enforcement) | South Africa |
| 30 | Mushayabasa, S. | 2020 | COVID-19 | ODE | Incidence/Prevalence | Contact/Transmission rate | Yes (Public perception of risk increases with confirmed cases and deaths) | No | No | No | Yes (Governmental actions like lockdowns and media announcements) | South Africa |
| 31 | Benneyan JC | 2020 | COVID-19 | ODE | Incidence/Prevalence | Contact/Transmission rate | Yes (uncertainty in infection and mortality rates) | Yes (effectiveness of testing, tracing, and isolation) | Yes (compliance to distancing precautions) | No | Yes (reopening plans and precaution awareness campaigns) | Theoretical |
| 32 | Teslya A | 2020 | COVID-19 | ODE | Incidence/Prevalence | Contact/Transmission rate | Yes (Disease awareness and perceived risk) | Yes (Efficacy of handwashing, mask-wearing, and social distancing) | No | No | Yes (Government-imposed social distancing and awareness spread) | Theoretical |
| 33 | Kain MP | 2020 | COVID-19 | ODE | Incidence/Prevalence | Contact/Transmission rate | No | No | No | No | Yes (social distancing orders) | United States |
| 34 | Kabir MH | 2020 | COVID-19 | ODE | Media/Awareness | Contact/Transmission rate | No | Yes (Social awareness control parameter) | No | No | Yes (Social distancing, wearing masks) | Bangladesh |
| 35 | Liu PY | 2020 | COVID-19 | ODE | Media/Awareness | Contact/Transmission rate | No | No | No | No | Yes (media reports) | Italy |
| 36 | Buonomo B | 2020 | COVID-19 | ODE | Media/Awareness | Contact/Transmission rate | Yes (awareness about the status of the disease in the community) | No | No | No | Yes (circulating information and rumours) | Italy |
| 37 | Adeniyi MO | 2020 | COVID-19 | ODE | Media/Awareness | Contact/Transmission rate | Yes (Awareness about disease and mode of transmission) | Yes (Precautionary measures like good hygiene) | No | No | Yes (Educational campaign and media propagation) | Theoretical |
| 38 | Kucharski AJ | 2020 | SARS-CoV-2 | ABM | Incidence/Prevalence | Contact/Transmission rate | Yes, the model considers symptomatic cases and their isolation. | No | No | No | Yes, app-based tracing and manual contact tracing are external triggers. | UK |
| 39 | Doncel J | 2020 | Theoretical | Game Theory | Imitation/Social Learning | Vaccination uptake | Yes (infection cost proportional to time spent infected enters the individual cost function) | Yes (explicit trade-off between vaccination cost and infection cost) | No | Yes (individual optimal vaccination depends on population state via mean field interactions) | No | Theoretical |
| 40 | Mancastroppa, M. | 2020 | Theoretical | Network Model | Incidence/Prevalence | Contact/Transmission rate | Yes (individuals avoid contact with infected nodes) | No | No | No | No | Theoretical |
| 41 | Arefin M.R | 2020 | Theoretical | ODE | Imitation/Social Learning | Vaccination uptake | No | Yes (aspiration levels depend on vaccine efficacy and cost) | No | Yes (imitation of peers) | No | Theoretical |
| 42 | Juher, D. | 2020 | Theoretical | ODE | Incidence/Prevalence | Contact/Transmission rate | Yes (risk perception is mentioned as a determinant of protective behavior) | No | No | No | No | Theoretical |
| 43 | Shanta, S.S. | 2020 | Theoretical | ODE | Media/Awareness | Contact/Transmission rate | No | No | No | No | Yes (media awareness programs) | Theoretical |
| 44 | Naik, P.A. | 2020 | Theoretical | ODE | Media/Awareness | Contact/Transmission rate | No | No | No | No | Yes (media coverage) | Theoretical |
| 45 | Gomez J | 2021 | COVID-19 | ABM | Exogenous/Policy | Contact/Transmission rate | No | No | No | No | Yes (social distancing policies) | Colombia |
| 46 | Pollmann, T.R. | 2021 | COVID-19 | ABM | Exogenous/Policy | Contact/Transmission rate | No | No | No | No | Yes (Digital contact tracing and testing protocols) | Theoretical |
| 47 | Ghosh S | 2021 | COVID-19 | ABM | Exogenous/Policy | Contact/Transmission rate | No | No | No | No | Yes (lockdown implementation) | Theoretical |
| 48 | Getz WM | 2021 | COVID-19 | ABM | Incidence/Prevalence | Contact/Transmission rate | Yes (adaptive contact rate influenced by prevalence) | No | No | No | No | Theoretical |
| 49 | Head JR | 2021 | COVID-19 | ABM | Incidence/Prevalence | Contact/Transmission rate | Yes (Risk of symptomatic illness and hospitalization) | Yes (Effectiveness of masks and social distancing) | No | No | Yes (School closure and reopening policies) | United States |
| 50 | Abueg, M. | 2021 | COVID-19 | ABM | Media/Awareness | Contact/Transmission rate | Yes (perceived risk of COVID-19 exposure) | No | No | No | Yes (exposure notifications and public health guidelines) | United States |
| 51 | Martcheva M | 2021 | COVID-19 | Game Theory | Incidence/Prevalence | Contact/Transmission rate | Yes (risk associated with not adopting social-distancing) | Yes (cost of strict social-distancing) | No | Yes (social norms to follow health directives) | No | Theoretical |
| 52 | Scabini, L.F.S. | 2021 | COVID-19 | Network Model | Exogenous/Policy | Contact/Transmission rate | No | No | No | No | Yes (isolation measures and lockdown policies) | Brazil |
| 53 | Nande, A. | 2021 | COVID-19 | Network Model | Exogenous/Policy | Contact/Transmission rate | No | No | Yes (Household size and occupation are considered structural barriers that increase individual risk and limit the efficacy of interventions) | No | No | Theoretical |
| 54 | Hill, E.M. et al. | 2021 | COVID-19 | Network Model | Exogenous/Policy | Contact/Transmission rate | No | No | Yes (Room isolation and rehousing constraints) | No | Yes (Testing and isolation policies) | United Kingdom |
| 55 | Zhao X | 2021 | COVID-19 | Network Model | Media/Awareness | Vaccination uptake | Yes (individuals become aware and take protective measures) | No | No | No | Yes (information from social networks and media) | China |
| 56 | Eksin C | 2021 | COVID-19 | Network Model | Incidence/Prevalence | Contact/Transmission rate | Yes (individuals reduce contacts based on cumulative cases) | No | No | No | Yes (awareness from neighboring localities) | Theoretical |
| 57 | Zhou, Y. | 2021 | COVID-19 | ODE | Exogenous/Policy | Contact/Transmission rate | No | No | No | No | Yes (government isolation measures) | China |
| 58 | AlJohani, N.I. | 2021 | COVID-19 | ODE | Exogenous/Policy | Contact/Transmission rate | No | No | No | No | Yes (government-imposed social distancing and self-isolation measures) | Saudi Arabia |
| 59 | Epstein, R. | 2021 | COVID-19 | ODE | Exogenous/Policy | Contact/Transmission rate | No | No | Yes (noncompliance and test device limitations) | Yes (pressure from friends, colleagues, and relatives) | Yes (media and government announcements) | Theoretical |
| 60 | Batabyal S | 2021 | COVID-19 | ODE | Exogenous/Policy | Contact/Transmission rate | No | No | No | No | Yes (lockdown, quarantine, social distancing) | Theoretical |
| 61 | Baba, B.A. | 2021 | COVID-19 | ODE | Exogenous/Policy | Contact/Transmission rate | No | Yes (control costs for awareness, lockdown, quarantine, and treatment explicitly | No | No | Yes (awareness campaigns, lockdowns, quarantine, and treatment | Theoretical |
| | | | | | | | | modeled in the objective function) | | | implemented as time-dependent controls) | |
| 62 | van Bunnik BAD | 2021 | COVID-19 | ODE | Exogenous/ Policy | Contact/Transmission rate | No | No | No | No | Yes (government advice and shielding recommendations) | United Kingdom |
| 63 | Kraay ANM | 2021 | COVID-19 | ODE | Exogenous/ Policy | Contact/Transmission rate | No | No | No | No | Yes (Contact reductions are triggered by specific dates corresponding to the enactment and lifting of stay-at-home orders and school reopenings) | United States |
| 64 | Raina MacIntyre, C. | 2021 | COVID-19 | ODE | Exogenous/ Policy | Contact/Transmission rate | No | No | No | No | Yes (mask mandates and recommendations) | United States |
| 65 | Skrip | 2021 | COVID-19 | ODE | Exogenous/ Policy | Contact/Transmission rate | No | No | Yes (Constraints around food, water, physical space, and sanitation are modeled as barriers to adopting home isolation) | No | Yes (Notification via Rapid Assessment Systems (RAS) is mentioned as a trigger for deployment of resources) | Liberia |
| 66 | Pinto Neto O | 2021 | COVID-19 | ODE | Incidence/Prevalence | Contact/Transmission rate | Yes (R_t greater than 1 indicates ongoing transmission) | Yes (Effectiveness of social distancing and protection measures) | No | No | Yes (Government mandates and public health guidelines) | Brazil |
| 67 | IHME COVID-19 Forecasting Team. | 2021 | COVID-19 | ODE | Incidence/Prevalence | Contact/Transmission rate | Yes (threshold of 8 deaths per million triggers policy changes) | Yes (effectiveness of mask use in reducing transmission) | No | No | Yes (state-level mandates and policy changes) | United States |
| 68 | Hanthanan Arachchilage, K. | 2021 | COVID-19 | ODE | Incidence/Prevalence | Contact/Transmission rate | Yes (fear-based model leading to mask-wearing) | No | No | No | No | United States |
| 69 | Bai, L. | 2021 | COVID-19 | ODE | Incidence/Prevalence | Contact/Transmission rate | Yes (risk of resurgence and imported cases) | Yes (efficacy of mask usage) | No | No | Yes (government control measures and guidelines) | China |
| 70 | Makhoul | 2021 | COVID-19 | ODE | Incidence/Prevalence | Contact/Transmission rate | Yes (Implicitly modeled as 'behavior compensation' where perception of protection post-vaccination leads to increased contact) | No | No | No | No | China |
| 71 | Di Domenico L | 2021 | COVID-19 | ODE | Incidence/Prevalence | Contact/Transmission rate | Yes (fear of contracting COVID-19) | No | No | No | Yes (media and government announcements) | France |
| 72 | Chan TL | 2021 | COVID-19 | ODE | Incidence/Prevalence | Contact/Transmission rate | Yes (perceived risk of infection) | Yes (perceived cost after infection) | No | No | Yes (information delay) | Hong Kong |
| 73 | Pillonetto G | 2021 | COVID-19 | ODE | Incidence/Prevalence | Contact/Transmission rate | Yes (people's growing awareness of infection risk) | No | No | No | Yes (lockdown restrictions) | Italy |
| 74 | Park SN | 2021 | COVID-19 | ODE | Incidence/Prevalence | Contact/Transmission rate | Yes (fear of infection) | No | No | No | Yes (news media reports) | Korea |
| 75 | Keeling MJ | 2021 | COVID-19 | ODE | Incidence/Prevalence | Contact/Transmission rate | Yes (awareness or fear of the spread of the virus) | No | No | No | Yes (government interventions and public efforts) | Korea |
| 76 | Ghosh, I. | 2021 | COVID-19 | ODE | Incidence/Prevalence | Contact/Transmission rate | Yes (Transmission rate and learning factor related to awareness of susceptibles are crucial) | No | No | No | Yes (Awareness through media and prosocial awareness effects) | Multi-countries |
| 77 | Bulut, H. | 2021 | COVID-19 | ODE | Incidence/Prevalence | Contact/Transmission rate | Yes (perceptions regarding the impending danger) | Yes (perceptions regarding the efficacy of the measures) | No | No | No | Multi-countries |
| 78 | Gozzi N | 2021 | COVID-19 | ODE | Incidence/Prevalence | Contact/Transmission rate | Yes (risk perception influenced by number of deaths) | No | No | No | Yes (media and communication of risk) | Multi-countries |
| 79 | Lee J | 2021 | COVID-19 | ODE | Incidence/Prevalence | Contact/Transmission rate | Yes (fear of disease proportional to daily confirmed cases) | No | No | No | Yes (government policies such as social distancing and lockdown) | Multi-countries |
| 80 | Eastman B | 2021 | COVID-19 | ODE | Incidence/Prevalence | Contact/Transmission rate | Yes (The model considers public response driven by oscillations in infection numbers, implying a reaction to perceived threat) | No | No | No | Yes (The public response function can be triggered by infection numbers reaching certain levels, acting as a cue) | Canada |
| 81 | Fields | 2021 | COVID-19 | ODE | Incidence/Prevalence | Contact/Transmission rate | No | No | No | No | Yes (Model includes a dampening variable, q, that represents the decrease in base contacts due to lockdown and social-distancing measures, with changes in q calibrated to specific dates of policy enactment) | Canada |
| 82 | Jentsch P.C | 2021 | COVID-19 | ODE | Incidence/Prevalence | Contact/Transmission rate | Yes (Population adherence to non-pharmaceutical interventions responds to case incidence) | No | No | Yes (Population adherence to non-pharmaceutical interventions) | No | Canada |
| 83 | Gatti, N. | 2021 | COVID-19 | ODE | Incidence/Prevalence | Contact/Transmission rate | Yes (contact rates decrease endogenously as a function of age- | No | No | No | No | Switzerland |
| | | | | | | | specific and aggregate COVID-19 death rates) | | | | | |
| 84 | Espinoza B | 2021 | COVID-19 | ODE | Incidence/Prevalence | Contact/Transmission rate | Yes (risk misperception of asymptomatic individuals) | Yes (trade-off between benefits of contact and infection risk) | No | No | No | Theoretical |
| 85 | Barzon, G. | 2021 | COVID-19 | ODE | Incidence/Prevalence | Contact/Transmission rate | Yes (awareness of potential danger of the virus) | No | Yes (impossibility for a fraction of the population to enter the hidden state) | No | Yes (cancellation of public events, local lockdowns) | Theoretical |
| 86 | Makris M. | 2021 | COVID-19 | ODE | Incidence/Prevalence | Contact/Transmission rate | Yes (Different infection-induced fatality rates) | Yes (Cost of social distancing and infection) | Yes (Exogenous bounds on social distancing) | No | No | United Kingdom |
| 87 | Duran-Olivencia, M.A. | 2021 | COVID-19 | ODE | Incidence/Prevalence | Contact/Transmission rate | Yes (social awareness and preventive measures) | No | No | No | Yes (government announcements and policies) | United Kingdom |
| 88 | Zu J | 2021 | COVID-19 | ODE | Incidence/Prevalence | Contact/Transmission rate | Yes (risk of COVID-19 infection) | Yes (effectiveness of face masks and social distancing) | No | No | Yes (government mandates and public health recommendations) | United States |
| 89 | Gros | 2021 | COVID-19 | ODE | Incidence/Prevalence | Contact/Transmission rate | Yes (behavioral and policy response triggered by cumulative or current case counts) | | Yes (social distancing and non-pharmaceutical interventions modeled as constraints reducing transmission) | | Yes (responses driven by reported case counts and epidemic history, including testing and policy signals) | Multiple countries |
| 90 | Nakamura, G. | 2021 | COVID-19 | ODE | Incidence/Prevalence | Vaccination uptake | Yes (vaccination awareness related to number of infective) | No | No | No | No | Theoretical |
| 91 | Bhattacharyya, R. | 2021 | COVID-19 | ODE | Media/Awareness | Contact/Transmission rate | No | Yes (Efficacy of social awareness and lockdowns) | No | No | Yes (Government campaigns and media awareness) | India |
| 92 | Tiwari PK | 2021 | COVID-19 | ODE | Media/Awareness | Contact/Transmission rate | No | Yes (Efficacy of information campaigns) | No | No | Yes (Global information campaigns) | India |
| 93 | Lacitignola, D. | 2021 | COVID-19 | ODE | Media/Awareness | Contact/Transmission rate | Yes (increased awareness pushes susceptibles towards social distancing and isolation) | No | No | No | Yes – awareness U(t) is driven by infections via implementation processes in the original SEIR model and treated as a dynamic control input in the Z-control model | Italy |
| 94 | Lacitignola, D. | 2021 | COVID-19 | ODE | Media/Awareness | Contact/Transmission rate | No | No | No | No | Yes – awareness variable m(t) increases when infections rise, reflecting media-driven information | Italy |
| 95 | Al-Tameemi, A.K.S. | 2021 | COVID-19 | ODE | Media/Awareness | Contact/Transmission rate | No | No | No | No | Yes – media alert term b(I) decreases contact rate as infections rise | Theoretical |
| 96 | Dwomoh, D. | 2021 | COVID-19 | ODE | Media/Awareness | Contact/Transmission rate | Yes (intensity of media coverage depends on the proportion of cases and deaths in the population, influencing perceived risk) | Yes (behavioral change parameter r explicitly models adherence to prevention guidelines (mask use, distancing, hygiene) reducing transmission) | No | No | Yes (intensive media coverage and government communication trigger increased adherence to prevention measures) | Ghana |
| 97 | Iboi E | 2021 | COVID-19 | ODE | Media/Awareness | Contact/Transmission rate | Yes (loss of willingness would increase mortality) | Yes (education efficacy in preventing infection) | No | No | Yes (public health education program) | United States |
| 98 | Li T | 2021 | COVID-19 | ODE | Media/Awareness | Contact/Transmission rate | Yes (risk perception) | No | No | No | Yes (media coverage) | China |
| 99 | Chang X | 2021 | COVID-19 | ODE | Media/Awareness | Contact/Transmission rate | Yes (negative emotions generated by confirmed cases) | No | No | No | Yes (media coverage and policies and regulations information) | China |
| 100 | Buonomo B | 2021 | COVID-19 | ODE | Media/Awareness | Vaccination uptake | Yes (perceived risk of COVID-19) | Yes (perceived efficacy of vaccination) | No | No | Yes (media coverage and public information) | Italy |
| 101 | Saito R | 2021 | Hepatitis A | ODE | Media/Awareness | Contact/Transmission rate | Yes – increased awareness of epidemic risk among MSM is inferred to reduce risky behaviors | No | No | No | Yes (campaign-based interventions (pamphlets and web articles) act as external informational triggers reducing transmission) | Japan |
| 102 | Jing, W. | 2021 | Syphilis | ODE | Media/Awareness | Contact/Transmission rate | Yes (public awareness (proxied by Baidu Index) reflects perceived risk and reduces infection probability) | Yes (protective behaviors (Ce) and condom-use compliance (?) explicitly reduce transmission) | No | No | Yes (increased media attention and public health education after 2013 trigger changes in protective behaviors) | China |
| 103 | Sajjadi, S. | 2021 | Theoretical | ABM | Incidence/Prevalence | Contact/Transmission rate | No | No | Yes (agents avoid physical barriers and other agents) | No | No | Theoretical |
| 104 | Du E | 2021 | Theoretical | ABM | Media/Awareness | Contact/Transmission rate | Yes (agents form a continuous opinion variable representing perceived infection risk based on information sources) | Yes (adoption of preventive behaviors is modeled via a risk-threshold decision reflecting perceived benefit of prevention) | No | Yes (opinions are influenced by social media interactions and neighbor observations) | Yes (global information releases and social media signals trigger opinion updating) | Theoretical |
| 105 | Kauk J | 2021 | Theoretical | Game Theory | Game-Theoretic | Vaccination uptake | Yes (individuals' perception of self-protective measures) | Yes (vaccine efficacy and cost) | Yes (cost of self-defense measures) | No | No | Theoretical |
| 106 | Huang, H. | 2021 | Theoretical | Network Model | Incidence/Prevalence | Contact/Transmission rate | Yes (risk perception from infected neighbors) | No | No | No | No | Theoretical |
| 107 | Amaral, M.A. | 2021 | Theoretical | ODE | Game-Theoretic | Contact/Transmission rate | Yes (perceived disease risk) | Yes (perceived cost of quarantine) | No | No | No | Theoretical |
| 108 | Epstein JM | 2021 | Theoretical | ODE | Incidence/Prevalence | Vaccination uptake | Yes (fear of disease and vaccine) | No | No | No | No | Theoretical |
| 109 | Ogunmiloro, O.M. | 2021 | Theoretical | ODE | Media/Awareness | Contact/Transmission rate | No | No | No | No | Yes (media awareness programs) | Theoretical |
| 110 | Baloba, E.B. | 2022 | Anthrax | ODE | Incidence/Prevalence | Contact/Transmission rate | Yes (behavioral change in humans is modeled as increasing with the number of infected individuals, reducing effective contacts with infected animals and contaminated environments) | Yes (positive behavioral change (e.g., awareness, reduced contact, improved hygiene) is explicitly modeled as lowering infection risk and increasing recovery-related outcomes) | No | No | Yes (mass awareness interventions and outbreak severity implicitly trigger behavioral change through the behavioral parameter in the incidence function) | Ghana |
| 111 | Bisanzio, D. | 2022 | COVID-19 | ABM | Exogenous/Policy | Contact/Transmission rate | No | No | No | No | Yes (government mandates and policies) | Saudi Arabia |
| 112 | d'Andrea V | 2022 | COVID-19 | ABM | Incidence/Prevalence | Contact/Transmission rate | Yes (risk perception) | No | No | No | No | Theoretical |
| 113 | Kordonis, I. | 2022 | COVID-19 | Game Theory | Game-Theoretic | Contact/Transmission rate | Yes (vulnerability of the player) | Yes (utility from interaction and disutility from infection) | Yes (minimum social contacts needed and government-imposed maximum) | No | No | Theoretical |
| 114 | Tang B | 2022 | COVID-19 | Game Theory | Game-Theoretic | Contact/Transmission rate | Yes (perceived risk of infection) | No | No | Yes (imitation dynamics) | Yes (self-awareness and NPIs) | Theoretical |
| 115 | Ferguson EA | 2022 | COVID-19 | ODE | Exogenous/Policy | Contact/Transmission rate | No | No | No | No | Yes (media campaigns and community support teams) | Bangladesh |
| 116 | Soares, A.L.O. | 2022 | COVID-19 | ODE | Exogenous/Policy | Contact/Transmission rate | No | No | No | No | Yes (government campaigns for social isolation) | Brazil |
| 117 | Yuan P | 2022 | COVID-19 | ODE | Exogenous/Policy | Contact/Transmission rate | No | No | No | No | Yes (Stay-at-home policy implementation) | Canada |
| 118 | Shah PV. | 2022 | COVID-19 | ODE | Exogenous/Policy | Contact/Transmission rate | No | No | No | No | Yes (government lockdown and social distancing measures) | India |
| 119 | Hezam IM. | 2022 | COVID-19 | ODE | Exogenous/Policy | Contact/Transmission rate | No | Yes (intervention strategies (social distancing, quarantine, testing, personal protection, insecticides) are | Yes (feasibility constraints and resource limitations (e.g., bounds on control variables, cost minimization, limited testing and | No | Yes (government-imposed interventions (social distancing, quarantine, testing availability, personal protection campaigns, | Yemen |
| | | | | | | | | explicitly modeled with associated costs in an optimal control framework) | quarantine capacity) are explicitly incorporated) | | insecticide use) act as external triggers modifying transmission dynamics) | |
| 120 | Armaou A | 2022 | COVID-19 | ODE | Exogenous/Policy | Contact/Transmission rate | No | Yes (reduction in transmission is explicitly linked to cautious behaviors and social distancing effectiveness within the transmission rate formulation) | Yes (limits on hospital bed capacity and feasibility constraints on activity curtailment are explicitly imposed in the control formulation) | No | Yes (government-imposed social distancing policies and public awareness campaigns act as external policy triggers modifying mobility and contact patterns) | Italy |
| 121 | Mamun-Ur-Rashid Khan, M.M.-U.-R. | 2022 | COVID-19 | ODE | Game-Theoretic | Contact/Transmission rate | Yes (individual self-quarantine decisions depend on perceived infection risk via the prevalence of visible infected individuals) | Yes (explicit costs and benefits of self-quarantine and forced quarantine are modeled in the behavior dynamics) | Yes (economic costs of self-quarantine for individuals and budget constraints for government-enforced quarantine are explicitly modeled) | No | Yes (government-enforced quarantine acts as an external trigger) | Theoretical |
| 122 | Agusto FB | 2022 | COVID-19 | ODE | Imitation/Social Learning | Contact/Transmission rate | Yes (perceived risk of infection) | Yes (perceived burden of infecting others vs. cost of self-isolation) | Yes (quarantine violation due to fatigue or material needs) | Yes (imitation dynamics/social learning) | Yes (publicly available information on COVID-19 death rates) | Theoretical |
| 123 | Safaie N | 2022 | COVID-19 | ODE | Incidence/Prevalence | Contact/Transmission rate | Yes ( perceived risk is represented through variables such as fraction infected and public thinking pressure that increase protective behavior and vaccine demand) | Yes (vaccination effectiveness, booster dose impact, and adherence to health protocols are explicitly modeled as reducing R0 and transmission) | Yes (economic strain, hospital capacity, vaccine availability, and access constraints are explicitly incorporated) | No | Yes (government policies, media communication, vaccination campaigns, and restriction changes act as external triggers affecting behavior) | Iran |
| 124 | Du Z | 2022 | COVID-19 | ODE | Incidence/Prevalence | Contact/Transmission rate | Yes (risk perception of infection) | No | No | No | Yes (public reports of daily cases) | Hong Kong |
| 125 | Santra, P.K. | 2022 | COVID-19 | ODE | Incidence/Prevalence | Contact/Transmission rate | Yes (psychological effect on human population) | No | No | No | No | India |
| 126 | Fierro A | 2022 | COVID-19 | ODE | Incidence/Prevalence | Contact/Transmission rate | Yes (risk perception during rising phases of the epidemic) | No | No | No | No | Italy |
| 127 | Liu S | 2022 | COVID-19 | ODE | Incidence/Prevalence | Contact/Transmission rate | Yes (People changed travel behavior based on information about COVID-19 cases) | No | Yes (Travel restrictions and stay-at-home requests) | No | Yes (Government announcements and state of emergency) | Japan |
| 128 | Althobaity, Y. | 2022 | COVID-19 | ODE | Incidence/Prevalence | Contact/Transmission rate | Yes (behavioural transitions back to adherence are driven by recent deaths per | Yes ( vaccination efficacy and continued NPI adherence are explicitly modeled as | No | No | No | Multi-countries |
| | | | | | | | 100,000, representing perceived severity and risk) | reducing infection risk and mortality) | | | | |
| 129 | Cordova-Lepe F | 2022 | COVID-19 | ODE | Incidence/Prevalence | Contact/Transmission rate | Yes (reaction rate is an increasing function of the number of infectious individuals, representing fear and perceived severity) | Yes (mitigation actions (social distancing, mask use, hygiene) reduce transmission through a reaction term lowering the transmission rate) | No | No | No | Multi-countries |
| 130 | Teslya, A. | 2022 | COVID-19 | ODE | Incidence/Prevalence | Contact/Transmission rate | Yes (Perceived susceptibility to infection) | No | No | No | Yes (Media and health authorities information) | Theoretical |
| 131 | Espinoza B | 2022 | COVID-19 | ODE | Incidence/Prevalence | Contact/Transmission rate | Yes (risk perceptions) | No | Yes (economic stress, lack of trust) | No | No | Theoretical |
| 132 | Canga, A. | 2022 | COVID-19 | ODE | Incidence/Prevalence | Contact/Transmission rate | Yes (Individuals react to the risk of disease transmission) | Yes (Effectiveness of self-protective measures) | No | No | Yes (Dissemination of information through media and other channels) | Theoretical |
| 133 | Musa SS | 2022 | COVID-19 | ODE | Incidence/Prevalence | Contact/Transmission rate | Yes (Human protective behavior reaction to recent deaths) | No | No | No | Yes (Nonpharmaceutical interventions and vaccination) | Multi-countries |
| 134 | Lau MSY | 2022 | COVID-19 | ODE | Incidence/Prevalence | Contact/Transmission rate | No | Yes (Adherence to facemask wearing) | No | No | Yes (Shelter-in-place order) | United States |
| 135 | Newcomb K | 2022 | COVID-19 | ODE | Incidence/Prevalence | Contact/Transmission rate | Yes (concern about transmission dynamics and infection resurgences) | Yes (evaluation of social interventions like social distancing and contact tracing) | No | No | Yes (policy decisions and phased lockdown release) | United States |
| 136 | Duarte, N. | 2022 | COVID-19 | ODE | Incidence/Prevalence | Quarantine adherence | Yes (Users notified of potential infection) | Yes (Adherence to quarantine recommendations) | No | No | Yes (Notification from wearable sensors) | Canada |
| 137 | Rai, R.K. | 2022 | COVID-19 | ODE | Media/Awareness | Contact/Transmission rate | No | Yes (Adoption of practices like wearing face masks, social distancing) | No | No | Yes (Social media advertisements) | India |
| 138 | Tripathi A | 2022 | COVID-19 | ODE | Media/Awareness | Contact/Transmission rate | No | No | No | No | Yes (media awareness campaigns) | India |
| 139 | Guo J | 2022 | COVID-19 | ODE | Media/Awareness | Contact/Transmission rate | Yes (media reporting Mt increases public risk perception, which reduces contact rates and increases quarantine) | Yes (effectiveness of preventive actions is explicitly modeled via reduced contact rate and increased quarantine efficiency driven by media intensity) | No | No | Yes (daily media reporting acts as an external informational trigger modifying behavior-related parameters) | China |
| 140 | Kumar S | 2022 | COVID-19 | ODE | Media/Awareness | Contact/Transmission rate | No | No | No | No | Yes (social media influence) | Theoretical |
| 141 | Abbas W | 2022 | COVID-19 | ODE | Media/Awareness | Contact/Transmission rate | Yes (fear of disease and awareness driven by high numbers of active cases move individuals from susceptible S to behavior-changed susceptible, Sf) | Yes (behavior-changed individuals have a reduced transmission rate, representing perceived effectiveness of preventive measures) | No | No | Yes (strengthening of preventive measures and public health guidelines in response to rising cases triggers behavioral change) | Multi-countries |
| 142 | Lopez-Cruz, R. | 2022 | COVID-19 | ODE | Media/Awareness | Quarantine adherence | Yes (the information index M depends on the infected population and feeds back into individual decisions to isolate) | Yes (behavioral change operates through an information-dependent isolation function p(M) representing perceived effectiveness of nonpharmaceutical interventions) | No | No | Yes (the information index (information/feedback about the epidemic and interventions) acts as an external informational trigger modifying isolation behavior) | Theoretical |
| 143 | Zuo C | 2022 | COVID-19 | ODE | Media/Awareness | Vaccination uptake | Yes (individuals perceive risk from global epidemic information) | Yes (individuals evaluate the efficacy of vaccination) | No | Yes (influence from aware neighbors) | Yes (media and government announcements) | China |
| 144 | Baloba, E.B. | 2022 | Ebola | ODE | Media/Awareness | Contact/Transmission rate | Yes (awareness of EVD symptoms and transmission) | No | No | No | Yes (media campaigns and public sensitization) | Multi-countries |
| 145 | Zhang H | 2022 | Influenza | ABM | Incidence/Prevalence | Contact/Transmission rate | No | Yes (effectiveness of vaccination (reduced susceptibility), mask-wearing (reduced transmission probability), and home-quarantine (reduced contacts) are explicitly parameterized) | Yes (limited compliance rates for vaccination, mask-wearing, and home-quarantine are explicitly modeled) | No | Yes (intervention measures (vaccination programs, mask-wearing, home-quarantine) are imposed as external control strategies evaluated under different compliance scenarios) | China |
| 146 | Panovska-Griffiths, J. | 2022 | Monkeypox | Network Model | Incidence/Prevalence | Contact/Transmission rate | Yes (perceived risk level) | No | No | No | Yes (media and social media activity) | United States |
| 147 | Yuan P | 2022 | Monkeypox | ODE | Exogenous/Policy | Vaccination uptake | No | Yes (Effectiveness of vaccination) | No | No | Yes (Testing and isolation policies) | Theoretical |
| 148 | Azizi, A. | 2022 | Theoretical | ABM | Incidence/Prevalence | Contact/Transmission rate | No | No | No | No | Yes (self-regulated social distancing triggered by symptomatic neighbors) | Theoretical |
| 149 | Amini, H. | 2022 | Theoretical | Network Model | Game-Theoretic | Contact/Transmission rate | Yes (probability of contracting the disease) | Yes (utility from social activity and cost of infection) | No | No | No | Theoretical |
| 150 | Horstmeyer L | 2022 | Theoretical | Network Model | Incidence/Prevalence | Contact/Transmission rate | Yes (nodes/agents avoid infected individuals due to the risk of acquiring the disease) | No | No | No | No | Theoretical |
| 151 | Buonomo B | 2022 | Theoretical | ODE | Incidence/Prevalence | Contact/Transmission rate | Yes (through information index related to disease prevalence) | No | No | No | Yes (through information index related to vaccination rollout) | Theoretical |
| 152 | Baba, I.A. | 2022 | Theoretical | ODE | Media/Awareness | Vaccination uptake | No | No | No | No | Yes (awareness increases transitions from anti-vaccine to pro-vaccine compartment) | Theoretical |
| 153 | Roosa K | 2022 | Zika | ODE | Incidence/Prevalence | Contact/Transmission rate | Yes (concern for disease transmission increases with the prevalence of infected humans and motivates uptake of personal protection) | Yes (personal protection is modeled as reducing transmission by a fixed efficacy (?), representing perceived effectiveness of protective actions) | No | No | No | Theoretical |
| 154 | Saad-Roy C.M | 2023 | SARS-CoV-2 | Game Theory | Game-Theoretic | Contact/Transmission rate | Yes (based on infection levels and perceived risk) | Yes (based on the cost of adhering to the NPI) | No | Yes (modeled through social learning) | No | Theoretical |
| 155 | Zuo C | 2023 | COVID-19 | ODE | Game-Theoretic | Vaccination uptake | Yes (individuals face a better trade-off in determining whether to get vaccinated based on the number of infectious individuals) | Yes (benefit-cost analysis includes perceived efficacy and cost of vaccination) | No | Yes (free-riding and altruism are considered in vaccination decisions) | Yes (authoritative media and global epidemic information influence vaccination decisions) | Theoretical |
| 156 | Yedomonhan E | 2023 | COVID-19 | ODE | Imitation/Social Learning | Contact/Transmission rate | Yes (individuals' perceived risk of COVID-19 influences their prophylactic opinion class, which directly modifies susceptibility to infection) | Yes (stronger prophylactic opinions correspond to higher perceived effectiveness of measures (masking, distancing, hygiene), modeled as reduced transmission parameters) | No | Yes (opinion changes occur through interpersonal influence and opinion amplification mechanisms within the susceptible population) | Yes (prevalence of infection and media/government information indirectly trigger opinion shifts by modifying influence functions tied to infection levels) | Multi-countries |
| 157 | Giaghedu M | 2023 | COVID-19 | ODE | Incidence/Prevalence | Contact/Transmission rate | Yes (agents reduce consumption and hours worked to lower infection probability) | No | No | No | No | Italy |
| 158 | Kolawole, M.K. | 2023 | COVID-19 | ODE | Incidence/Prevalence | Contact/Transmission rate | No | No | Yes (physical constraints like curfews and social distancing) | No | No | Nigeria |
| 159 | Avusuglo WS | 2023 | COVID-19 | ODE | Incidence/Prevalence | Contact/Transmission rate | Yes (Community compliance rate) | Yes (Effectiveness of compliance rate) | No | No | Yes (Media campaigns) | Theoretical |
| 160 | Song P | 2023 | COVID-19 | ODE | Incidence/Prevalence | Contact/Transmission rate | Yes (depends on prevalence and changing rate of prevalence) | No | No | No | No | Canada |
| 161 | Sardar, T. | 2023 | COVID-19 | ODE | Media/Awareness | Contact/Transmission rate | No | No | No | No | Yes (Media awareness campaigns) | India |
| 162 | Capistran, M.A. | 2023 | COVID-19 | ODE | Media/Awareness | Contact/Transmission rate | Yes (awareness-based behavior reflects risk perception leading to changes in isolation and hygiene, modeled via the effective population fraction) | No | No | No | Yes (governmental control measures and public awareness during the early pandemic phase act as external triggers reducing effective contacts) | Multi-countries |
| 163 | El Kihal F | 2023 | COVID-19 | ODE | Media/Awareness | Contact/Transmission rate | Yes (media-induced fear and awareness influence individual susceptibility and contact behavior through stochastic transmission terms) | Yes (immunization (vaccination/awareness) is modeled as an active control reducing susceptibility and infection risk with explicit cost term) | No | No | Yes (media coverage, infodemics, and policy-driven immunization and closure strategies act as external triggers modifying behavior and controls) | Theoretical |
| 164 | Addai E | 2023 | Marburg virus | ODE | Media/Awareness | Contact/Transmission rate | Yes – awareness and public health education reduce susceptibility and perceived risk of infection | Yes – effectiveness of public health education and awareness is explicitly modeled as reducing infection risk | No | No | Yes – public health education and information dissemination campaigns act as external triggers modifying behavior | Theoretical |
| 165 | Hadley L | 2023 | Meningococcal disease | ODE | Exogenous/Policy | Contact/Transmission rate | No | No | Yes (reduced healthcare capacity and disrupted routine vaccination during the COVID-19 pandemic are explicitly imposed as constraints on vaccine uptake) | No | Yes (government-imposed COVID-19 social distancing measures and school closures act as external policy triggers reducing social contacts) | United Kingdom |
| 166 | Bragazzi, N.L. | 2023 | Monkeypox | ODE | Incidence/Prevalence | Contact/Transmission rate | Yes (Risk perception and disease knowledge) | No | No | No | Yes (Awareness campaigns and public health policies) | Canada |
| 167 | Brand SPC | 2023 | Monkeypox | ODE | Incidence/Prevalence | Contact/Transmission rate | Yes (awareness of Mpox symptoms and intention to reduce transmission risk) | No | No | No | Yes (public health emergency declaration by WHO) | United Kingdom |
| 168 | Achterberg, M.A. | 2023 | Theoretical | ODE | Incidence/Prevalence | Contact/Transmission rate | Yes (personal risk perception) | No | No | No | No | Theoretical |
| 169 | Harris MJ | 2023 | Theoretical | ODE | Incidence/Prevalence | Contact/Transmission rate | Yes (perceived risk of infection based on awareness of disease burden) | No | No | No | Yes (awareness of epidemic conditions triggers protective behavior) | Theoretical |
| 170 | Diaz-Infante, S. | 2023 | Theoretical | ODE | Incidence/Prevalence | Contact/Transmission rate | Yes (Risk perception influences behavior) | No | No | No | Yes (Traffic light system as external trigger) | Theoretical |
| 171 | Wanjala, H.M. | 2023 | Theoretical | ODE | Media/Awareness | Contact/Transmission rate | Yes (Perception by the public arising from media campaign) | Yes (Efficacy of face-mask and its compliant and consistent use) | Yes (Reinfection due to waning immunity and stochastic perturbations (infodemics, escapes) are explicitly incorporated as structural constraints) | No | Yes (Media awareness programs) | Theoretical |
| 172 | Agrawal, S. | 2023 | Theoretical | ODE | Media/Awareness | Contact/Transmission rate | No | No | No | No | Yes (media coverage) | Theoretical |
| 173 | Hu L | 2023 | Vector-borne | ODE | Media/Awareness | Contact/Transmission rate | No | No | No | No | Yes (media coverage) | Theoretical |
| 174 | Snellman JE | 2024 | COVID-19 | ABM | Media/Awareness | Contact/Transmission rate | Yes (Fear of catching the disease and perceived infection risk explicitly influence population agents' behavior and mobility) | Yes (Perceived effectiveness and costs of mitigation measures (restrictions, reduced economic activity, vaccination) are explicitly weighed by population and authority agents) | Yes (Economic constraints and trade-offs between health protection and economic activity are structurally embedded in the model) | Yes (Compliance with authorities and socially driven responsibility are modeled through agent-based valuation and interaction mechanisms) | Yes (Government restrictions, real-time infection information, vaccination rollout, and policy stringency act as external triggers modifying behavior) | Spain |
| 175 | Smirnova, A. | 2024 | COVID-19 | Network Model | Exogenous/Policy | Contact/Transmission rate | No | Yes (Effectiveness of mask-wearing) | No | No | Yes (Government guidelines) | Korea |
| 176 | Li T | 2024 | COVID-19 | ODE | Imitation/Social Learning | Contact/Transmission rate | Yes (perceived risk of infection) | Yes (cost of self-imposable prophylactic measures) | No | Yes (imitation process) | No | Theoretical |
| 177 | Are EB | 2024 | COVID-19 | ODE | Incidence/Prevalence | Contact/Transmission rate | Yes (Perceived risk of COVID-19 infection) | No | No | Yes (Vaccine homophily and social network influence) | No | Canada |
| 178 | Brankston, G. | 2024 | COVID-19 | ODE | Incidence/Prevalence | Contact/Transmission rate | Yes (individuals respond to increasing disease risk) | No | No | No | Yes (policy announcements and media coverage) | Canada |
| 179 | De Gaetano A | 2024 | COVID-19 | ODE | Incidence/Prevalence | Contact/Transmission rate | Yes (Perceived severity of disease) | No | No | No | No | Italy |
| 180 | Tulchinsky A | 2024 | COVID-19 | ODE | Incidence/Prevalence | Contact/Transmission rate | Yes (perceived risk based on hospitalizations) | No | No | No | No | United States |
| 181 | Kuwahara B | 2024 | COVID-19 | ODE | Incidence/Prevalence | Contact/Transmission rate | Yes,( the model considers infection risk influencing behavior) | Yes, the model includes assumptions about the efficacy of interventions. | No | No | No | Theoretical |
| 182 | Zhang, X.-S. | 2024 | COVID-19 | ODE | Media/Awareness | Contact/Transmission rate | Yes (voluntary behaviour change is driven by perceived epidemic severity, reflected through reductions in contact rates inferred from Google mobility data) | Yes (protective behaviours (mobility reduction, distancing) and vaccination are explicitly modeled as reducing transmission and susceptibility) | No | No | Yes (government NPIs (captured via Oxford Government Response Index) and public information reflected in mobility data act as external triggers modifying behaviour) | Laos |
| 183 | Zhao, L. | 2024 | COVID-19 | ODE | Media/Awareness | Vaccination uptake | Yes (infection prevalence and disease information increase awareness and influence individuals' decisions to adopt vaccination as a protective behavior) | Yes (vaccine efficacy against infection, severe disease, and death is explicitly modeled and affects individual vaccination decisions and epidemic outcomes) | Yes (vaccine hesitancy and distrust (modeled via information-dissemination states and transition rates) limit vaccination adoption despite availability) | Yes (information spreads through a network structure where individuals influence others' trust and vaccination behavior) | Yes (dissemination of vaccine and disease information through media and social networks acts as an external trigger for behavioral change) | Iceland |
| 184 | Kengne, J.N. | 2024 | Ebola | ODE | Incidence/Prevalence | Contact/Transmission rate | Yes (perceived risk increases with numbers of infected, deceased, and environmental contamination, explicitly driving behavior change via the nonlinear incidence function) | Yes (efficacy of behavior change (p) and speed of behavioral response (g) explicitly reduce transmission, representing perceived effectiveness of protective actions) | No | No | No | Congo |
| 185 | Kiemtore A. | 2024 | Hepatitis B | ODE | Media/Awareness | Contact/Transmission rate | No | Yes (vaccination, treatment of chronic carriers, and media awareness efficacy are explicitly parameterized as reducing)transmission and/or prevalence | Yes (vaccine immunization failure and waning and limited treatment coverage/efficacy are explicitly modeled constraints) | No | Yes (media awareness campaigns act as an external trigger reducing infection risk among susceptible adults ) | Burkina Faso |
| 186 | Buonomo B | 2024 | Meningitis | ODE | Media/Awareness | Vaccination uptake | Yes (perceived risks associated to the disease) | No | No | No | Yes (information index summarizing current and past disease trend) | Nigeria |
| 187 | Cordeiro R | 2024 | Monkeypox | ODE | Incidence/Prevalence | Contact/Transmission rate | Yes ( transmission dynamics depend on epidemic intensity, with behavioral sensitivity reflected through changes in sexual activity patterns and awareness during outbreak waves) | Yes (vaccination is explicitly modeled as reducing susceptibility, transmission contribution, and disease severity) | No | No | No | Portugal |
| 188 | Lin, Y.-C. | 2024 | Monkeypox | ODE | Incidence/Prevalence | Contact/Transmission rate | Yes (High-risk group awareness) | No | No | Yes (High-risk group awareness among MSM) | Yes (Awareness due to outbreak and vaccination campaigns) | United States |
| 189 | Pei H | 2024 | Theoretical | ODE | Imitation/Social Learning | Contact/Transmission rate | No | No | No | Yes (influence of social proof) | No | Theoretical |
| 190 | Adikari, A.U.S. | 2025 | COVID-19 | ODE | Exogenous/Policy | Contact/Transmission rate | No | Yes (vaccination and social distancing controls are explicitly modeled with efficacy and cost terms that reduce infection and disease burden) | Yes (economic and feasibility constraints are explicitly imposed through bounded control variables and cost weights in the optimal control formulation) | No | Yes (government-imposed interventions act as external control triggers modifying transmission dynamics) | Sri Lanka |
| 191 | Chen, K. | 2025 | COVID-19 | ODE | Incidence/Prevalence | Contact/Transmission rate | Yes (protection awareness) | Yes (protection efficiency) | No | No | Yes (media coverage) | China |
| 192 | Chen, X. | 2025 | COVID-19 | ODE | Incidence/Prevalence | Contact/Transmission rate | Yes (perception of risk and severity of disease) | No | No | No | Yes (policy-driven behavior modifications) | Theoretical |
| 193 | Wang H | 2025 | COVID-19 | ODE | Media/Awareness | Contact/Transmission rate | Yes (perception of potential risks) | Yes (self-regulated or policy-driven behavior change) | No | No | Yes (media campaigns) | China |
| 194 | Dayap, J.A. | 2025 | Dengue | ODE | Incidence/Prevalence | Contact/Transmission rate | Yes (the transition rate from non-adherent to adherent behavior is an explicit increasing function of community infection levels) | Yes (efficacy of self-protective measures (?) and vector control (induced mosquito death rate are explicitly modeled as reducing transmission) | No | No | No | Taiwan |
| 195 | De Reggi S | 2025 | Dengue | ODE | Incidence/Prevalence | Contact/Transmission rate | Yes (individuals adapt behavior based on prevalence information) | No | No | No | No | Theoretical |
| 196 | Andrawus, J. | 2025 | Diphtheria | ODE | Media/Awareness | Contact/Transmission rate | Yes (awareness parameters shift individuals from unaware to aware exposed states, reflecting | Yes – (awareness, surveillance, vaccination, isolation, and personal hygiene are explicitly modeled as effective measures | No | No | Yes (surveillance programs and awareness campaigns act as external informational triggers increasing detection, | Nigeria |
| | | | | | | | perceived susceptibility and risk) | reducing transmission and progression) | | | isolation, and preventive behavior) | |
| 197 | Ahmed KK | 2025 | Dysentery | ODE | Media/Awareness | Contact/Transmission rate | Yes (individuals move into an educated susceptible class with reduced infection risk, representing increased perceived susceptibility and risk awareness) | Yes (public awareness, hygiene/sanitation, and treatment controls are explicitly modeled as effective interventions reducing infection and disease burden) | No | No | Yes (public awareness campaigns and hygiene promotion are modeled as external control efforts that trigger behavior change) | Theoretical |
| 198 | Buonomo B | 2025 | Meningococcal meningitis | ODE | Media/Awareness | Vaccination uptake | Yes (perceived risks associated to meningitis) | No | No | No | Yes (information and rumours about the disease) | Nigeria |
| 199 | Onifade AA | 2025 | Monkeypox | ODE | Media/Awareness | Contact/Transmission rate | No | No | Yes (difficulty in self-isolation due to jobs) | No | Yes (public awareness campaigns) | United States |
| 200 | Chong NS | 2025 | Other | ODE | Incidence/Prevalence | Contact/Transmission rate | Yes (Behavioral change when the number of infected dogs increases) | No | No | No | No | Theoretical |
| 201 | Buonomo B | 2025 | Respiratory viruses | ODE | Media/Awareness | Contact/Transmission rate | Yes (perceived risk of infection) | No | No | No | Yes (information and rumours) | Theoretical |
| 202 | Li T | 2025 | Theoretical | Game Theory | Incidence/Prevalence | Contact/Transmission rate | Yes (individuals adjust behavior based on perceived infection risks) | No | No | No | Yes (media and social networks influence behavior) | Theoretical |
| 203 | Mazza, F. | 2025 | Theoretical | Network Model | Incidence/Prevalence | Contact/Transmission rate | Yes (increased susceptibility and infectivity for non-compliant individuals) | No | No | No | No | Italy |
| 204 | Bentkowski, P. | 2025 | Theoretical | Network Model | Incidence/Prevalence | Contact/Transmission rate | No | Yes (impact of mask wearing and vaccination) | No | No | No | Theoretical |
| 205 | Schnyder SK | 2025 | Theoretical | ODE | Game-Theoretic | Contact/Transmission rate | Yes (perceived infection cost) | Yes (cost of social distancing) | No | No | No | Theoretical |
| 206 | Chakraborty A | 2025 | Theoretical | ODE | Game-Theoretic | Vaccination uptake | Yes (perceived risks and benefits) | Yes (perceived efficacy/cost of intervention) | Yes (financial and logistical burden) | Yes (imitation dynamics) | Yes (awareness or willingness to get tested) | Theoretical |
| 207 | Li L | 2025 | Theoretical | ODE | Imitation/Social Learning | Contact/Transmission rate | Yes (individuals change behavior based on perceived risk of infection) | No | No | Yes, behavior change is modeled through imitation processes. | No | Theoretical |
| 208 | Chen W | 2025 | Theoretical | ODE | Incidence/Prevalence | Contact/Transmission rate | Yes (spontaneous avoidance behaviors) | No | No | Yes (collective avoidance behaviors) | No | Theoretical |
| 209 | Buonomo B | 2025 | Theoretical | ODE | Incidence/Prevalence | Contact/Transmission rate | Yes (individuals modify behavior in response to perceived risk of infection) | No | No | No | Yes (awareness-driven behavioral changes) | Theoretical |
| 210 | Bouka M | 2025 | Theoretical | ODE | Incidence/Prevalence | Contact/Transmission rate | Yes (based on delayed information related to personal risk) | No | No | No | Yes (information feedback delays) | Theoretical |
| 211 | Buonomo B | 2025 | Theoretical | ODE | Incidence/Prevalence | Contact/Transmission rate | Yes (perceived risks associated with the disease) | No | No | No | No | Theoretical |
| 212 | Centrone, F. | 2025 | Theoretical | ODE | Incidence/Prevalence | Vaccination uptake | Yes (information on disease mortality influences vaccination decisions) | Yes (rational exemption based on perceived risk of vaccination side-effects) | No | No | No | Theoretical |
| 213 | Seibel RL | 2025 | Theoretical | ODE | Incidence/Prevalence | Vaccination uptake | Yes (perceived risk of infection) | Yes (real-time intervention effectiveness) | No | No | Yes (real-time outbreak information) | Theoretical |
| 214 | Qin, W. | 2025 | Theoretical | ODE | Media/Awareness | Contact/Transmission rate | No | No | No | No | Yes (media communication measures) | Theoretical |
| 215 | Aldila D | 2025 | Tuberculosis | ODE | Media/Awareness | Contact/Transmission rate | No | Yes (Effectiveness of media campaigns and mask use) | Yes (Limited treatment capacity) | No | Yes (Media campaigns) | Indonesia |
| 216 | Nawaz, R. | 2025 | Tuberculosis | ODE | Media/Awareness | Treatment uptake | Yes (media awareness increases perceived risk and knowledge among infected migrants and seasonal farm workers, influencing progression to treatment) | Yes (media impact parameter explicitly increases preventive treatment-seeking behavior) | No | No | Yes (media awareness acts as an external informational trigger affecting treatment adoption) | Theoretical |